\documentclass[doublespacing]{elsart}
\usepackage{graphicx}
\usepackage{amssymb}
\usepackage{epsfig}
\usepackage{epsf}
\usepackage{amsmath}
\usepackage{subfigure}
\journal{Sensors and Actuators B}
\begin{document}

\begin{frontmatter}

\title{Current correlation functions for chemical sensors based on DNA decorated carbon nanotube}

\author{P. Poonam},
\author{N. Deo\corauthref{cor}}
\corauth[cor]{Corresponding author. Tel.:+91-11-27667725 ext. 1342;
fax: +91-11-27667061.} \ead{ndeo@physics.du.ac.in}
\address{Department of Physics and Astrophysics,
University of Delhi, Delhi 110007, India}

\begin{abstract}
The current characteristics of DNA decorated carbon nanotubes for
different gas odors are studied. A simple model of charge transfer
between the Gas-DNA-base complex and single wall carbon nanotube
(SWCN) is proposed to explain the current response for different
odors. The autocorrelation and two-point correlation functions are
calculated for the current sensitivity curves. These correlation
functions together with the current characteristics form
finger-prints for detection of the odor and DNA sequence.

\end{abstract}

\end{frontmatter}

\newpage

\section{Introduction}
\label{1.}

The discovery of carbon nanotubes (CNTs) \cite{1} has generated a
tremendous amount of interest and activity in basic research and
applied technologies. The unique properties of CNTs make them a
potentially ideal material for molecular sensing.  The simplest
nanotube is a single layer of graphite rolled into a seamless
cylinder and is called the single wall carbon nanotube (SWCN)
having diameters ranging from 1-2 nm \cite{2}. This unique
structure results in electronic and chemical properties that are
ideal for the direct electronic detection of chemical vapors
\cite{3}. The electronic properties of CNTs can be altered by
modifying their parent structure \cite{4} and this can be done by
doping them with certain molecules or polymers
\cite{5,6,7,8,9,10,11,12,13,14,15}. It is found that the
conductance of SWCNs changes in response to exposure to certain
molecules that undergo charge transfer upon adsorption
\cite{5,6,7,8,9,10,11,12,13,14,15}. In particular, CNTs can be
functionalized with single stranded deoxyribonucleic acid
(ss-DNA) or DNA nucleoside \cite{4,16,17,18,19,20}.
Functionalization of CNTs with DNA offers interesting prospects
in different fields including detection of chemical vapors,
solubilization in aqueous media, and nucleic acid sensing
\cite{4,17,18,19,20}.

This work presents a theoretical understanding of the current
characteristics of gas flow over DNA decorated CNTs \cite{17}. The
article \cite{17} studies the nanoscale chemical sensors, with
ss-DNA for chemical recognition and single walled carbon nanotube
field effect transistors (SWCN-FETs) for electronic readout. The
composites, SWCNs coated with ss-DNA, respond to various gas odors
that do not respond or cause a detectable conductivity change in
the bare devices (without DNA). As a result of functionalization
of SWCNs with DNA a change in the current is observed for each gas
odor.

These findings motivated us to analyze the ss-DNA-Gas/SWCN complex
system and to understand the reason for the observed change in the
current. Our analysis gives two main results: the first is the
proposal of a simple phenomenological model based on charge transfer
between the Gas-DNA-base complex and the SWCN, and the second is the
calculation of the correlation functions for the current
characteristic curves of Ref. \cite{17}. We find that the current
fluctuations (correlation functions) together with the current
characteristics form finger-prints for detection of the odor. More
importantly the characteristic pattern of the DNA sequence is
captured in these current correlation functions. Hence these gas
flow sensors may also be used as sequence detectors for DNA where
the pattern of correlation functions may be used as a benchmark for
the particular chemical signal encoded in a DNA sequence.

\section{Experimental details}
\label{2.}

SWCNs were synthesized by catalytic chemical vapor deposition
(CVD) of methane (2.50 sLm\footnote{Standard liters per minute.})
at ${\rm 900^o C}$ on a ${\rm SiO}_2$/Si substrate using iron
salt catalyst (${\rm {Fe(NO_3)_3.9H_2O}}$ dissolved in isopropyl
alcohol) \cite{17}. Hydrogen (0.320 sLm) and argon (0.600 sLm)
are allowed to flow through the furnace throughout the heating
and growth process. FETs were fabricated with Cr/Au source and
drain electrodes using e-beam lithography and the degenerately
doped Si substrate used as a backgate \cite{17,18,19}. The
resistance of the FETs was $100-500 ~{\rm k \Omega}$ \cite{19} in
the ``ON" state.

The two ss-DNA sequences chosen in the experiment \cite{17} are:

Sequence 1: $5^{\prime}$ GAG TCT GTG GAG GAG GTA GTC $3^{\prime}$

Sequence 2: $5^{\prime}$ CTT CTG TCT TGA TGT TTG TCA AAC
$3^{\prime}$.

These oligonucleotides were diluted in distilled water to make a
stock solution of $658\mu g/ml$ (sequence 1) and $728\mu g/ml$
(sequence 2). First the odor responses of bare devices were
measured, and then the devices were functionalized with a
particular ss-DNA sequence by applying a $500\mu m$ diameter drop
of the solution to the device for 45 min and then dried in a
nitrogen stream \cite{17,18,19}.

The adsorption of ss-DNA on SWCNs was characterized by Atomic
Force Microscopy (AFM). The AFM images \cite{17,18,19} of the
same SWCN before and after the application of DNA show a clear
increase in the tube diameter from $5.4\pm0.1$ nm to $7.2\pm0.2$
nm\footnote{The mentioned diameter is most likely for a small
bundle of nanotubes \cite{19} as the CVD process used in Refs.
[17-19] tends to produce nanotubes with a diameter from 1-2 nm.}
indicating formation of a nanoscale layer of ss-DNA on the SWCN
surface \cite{17,18,19}. Functionalization of SWCNs with ss-DNA
caused the threshold value of the gate voltage $V_g$ to decrease
by 3-4 V for measurable conduction and this corresponds to a
decrease in the hole density \cite{17,18,19}.

The sensor response for five different odors Methanol, Propionic
acid (PA), Trimethylamine (TMA), 2, 6 Dinitrotoluene (DNT) and
Dimethyl methylphosphonate (DMMP) was studied. Here the focus was
on odor induced changes in the current measured with bias voltage
$V_b=100$mV and $V_g=0$V \cite{17}. In this experiment, gas and
then air is alternately exposed to the device each for 50
seconds. The experiment is repeated for multiple odor exposures
\cite{17}.

The bare and polymer coated SWCNs respond to various gases
\cite{3,5,6,7,8,9,10,11,12,13,14,15} but some chemical species
interact weakly or not at all with the bare SWCNs \cite{17,18,19}.
For molecular sensing, it is necessary that the chemical species
get adsorbed on the device properly. When SWCNs are coated with
DNA the bases bind to SWCNs through vdW forces and by forces due
to their mutual polarization \cite{16,20}. We propose that the
chemical species get adsorbed on SWCNs through vdW forces and/or
mutual polarization between the chemical species and DNA-SWCN
complex (as in Refs. \cite{16,20}), Fig. 1. Hence the DNA
increases the binding affinity of the molecular species (odors)
to the device. The model described below is a phenomenological
model. The building of a microscopic (tight binding) model of
CNTs is an important task and needs future work.

\section{Model}
\label{3.}

Here we propose a model to explain the sensor response \cite{17}.
The model consists of a FET in which a p-type SWCN acts as an
electronic wire between two metal electrodes with an ss-DNA
sequence on it. For clarity of exposition the model illustrated
in Fig. 2 has a simpler geometry than the experiment.

\emph{DMMP and 2,6 DNT}: The current response of the bare device
is less than the experimental sensitivity ($\bigtriangleup I/I$
$\sim 1\%$; where I is the source-drain current) when exposed to
DMMP with estimated concentration 25 ppm \cite{17}. After coating
the same device with ss-DNA sequence 2, exposure to DMMP gives a
slight change (decrease) in the current, that is, the sensitivity
of the device is $\bigtriangleup I/I= -7\%$ \cite{17}. Due to the
application of ss-DNA layer, the binding affinity of DMMP to the
device increases through vdW forces and/or mutual polarization
\cite{20}, this results in an increase in sensor response. The
fluctuations in the current response are due to the structure of
DNA sequence.

To explain the adsorption of ss-DNA on SWCNs and sensor response
mechanism molecular dynamics simulations were done in
\cite{18,19}. Another mechanism which needs to be modified to
include the effect of the DNA can be a pulsating asymmetrical
ratchet model used to describe the flow induced voltage
(current)\cite{21}. This may be a future work. Here we propose a
simpler mechanism which includes the DNA to describe the current
characteristics.

The DNA bases interact with the SWCN in the same way as described
in Ref.\cite{20}. The interaction of DMMP with DNA-bases through
vdW forces and/or mutual polarization \cite{20} causes charge
redistribution and changes the polarization between the DNA and
the SWCN leading to a net charge transfer from the Gas-DNA-base
complex to the SWCN. These charges are some combinations of
$\alpha$, $\beta$, $\gamma$ and $\delta$ negative ions which are
arbitrary fractional charges transferred by each Gas-DNA-base
complex and depend on the DNA sequence and odor. The net charge
transfer causes the Fermi level of the SWCN to shift away from
the valence band \cite{5}, which results in hole depletion of the
p-type SWCN-FET sensor and reduces the current. Note that the
charge transfer from the Gas-DNA-base complex to the SWCN is
larger than the charge transfer from DNA-base to the SWCN as well
as from SWCN to the Air-DNA-base complex.

This process continues till all DMMP molecules interact with the
full DNA sequence causing a net charge transfer which leads to a
minimum value of the current. A similar mechanism occurs for
multiple exposures of the same odor and we get the same minimum
value of the current. Thus for each odor exposure the pattern
repeats itself indicating the stability of the sensor.

For the next 50 seconds, when air is exposed to the device it
replaces the gas molecules as shown in Fig. 2 and interacts with the
SWCN through DNA. This causes a net charge transfer from the SWCN to
oxygen \cite{11} which shifts the Fermi level closer to the valence
band. This increases the hole concentration and enhances the
current. This process continues until the maximum value of the
current $I=I_0$ is reached, where $I_0$ is the source-drain current
when exposed to air (no odor) \cite{17}. For DNT (40 ppm) the
explanation is the same and the sensitivity of the device is
$\bigtriangleup I/I= -14\%$ with sequence 1 \cite{17}.

\emph{TMA}: The bare device responds to TMA (20000 ppm)\cite{17},
because it interacts with the SWCN through vdW forces and/or
mutual polarization \cite{20}. The negative charge transfer from
TMA to the SWCN neutralizes the holes and reduces the carrier
concentration of the sensor and there is a decrease in the
current. The observed sensitivity is $\bigtriangleup I/I= -10\%$.
After the application of DNA sequence 2 the response is tripled,
$\bigtriangleup I/I= -30\%$ \cite{17}. The polarization between
the SWCN and TMA-DNA-base complex causes charge transfer from the
TMA-DNA-base complex to the SWCN. There is an additional charge
transfer due to direct interaction of TMA with the SWCN. Both
processes cause a net charge transfer which is far greater than
the charge transfer before the application of DNA. On air
exposure the current increases and reaches a maximum value
($I=I_0$).

\emph{PA}: The sensor response to PA (150 ppm) differs in sign and
magnitude from the response to other odors \cite{17}. The
interaction of PA with DNA sequence 1 causes a net charge
transfer from SWCN to the Gas-DNA-base complex therefore
increases the hole concentration and the current response. In
this case the sensitivity is $\bigtriangleup I/I= +17 \%$
\cite{17}. On air exposure the current decreases and reaches a
minimum value $I_0$ (maximum for other odors) as the charge
transfer from SWCN to the Air-DNA-base complex is small compared
to the charge transfer to the PA-DNA-base complex.

\emph{Methanol}: The bare device does not respond to Methanol (4000
ppm). Exposure to Methanol gives $\bigtriangleup I/I= -12\%$ with
sequence 1 and $\bigtriangleup I/I= -20\%$ with sequence 2,
\cite{17}. The explanation for the sensor response is the same as
given for other odors except PA.

The important aspects of investigation of a variety of sensors are
their sensitivity, selectivity, and stability \cite{22}. The above
explanation shows how functionalization of SWCNs with ss-DNA is
helpful in increasing the sensitivity of the ss-DNA/SWCN-FETs by
increasing the binding affinity of the gas molecules to the devices
as the bare devices do not respond to these odors except TMA. Table
1 of Ref. \cite{17} shows measured responses of 5-10 devices to gas
odors. The interaction of different odors with SWCNs gives different
charge transfers leading to distinct patterns of the current
characteristics for each odor. This indicates selective recognition
of each odor by the ss-DNA/SWCN-FET sensors.

For each odor a constant sensor response is maintained without any
need for sensor refreshing through at least 50 gas exposure cycles
\cite{17}. The sensor response and recovery time remained the same
to within 5\%. The variation in odor responses for different
devices is small e.g. the sensitivity variation for TMA (sequence
2) is $\bigtriangleup I/I= -30 \pm 2\%$ measured with 5-10 devices
\cite{17}. This indicates excellent reproducibility and hence the
stability of the sensors.

If the device is exposed to a mixture of gases then the resulting
current characteristics will be different from the
characteristics observed for each individual gas odor. This is
because the sensing material does not respond to a particular
odor only, but to a mixture of gas odors. In this case it is
difficult to distinguish different odors from the resulting
characteristics. This indicates that the selectivity of a sensor
gets affected by a mixture of gases.

\section{Explanation for the difference in the odor response
characteristics of the sensors for two different sequences}
\label{4.}

We observe, in the Table 1 \cite{17}, that there is a larger
decrease in the current when TMA/Methanol interacts with DNA
sequence 2 than with sequence 1. This indicates that the net
charge transfer from the Gas-DNA-base complex to the SWCN is more
when TMA/Methanol interacts with sequence 2. We give here a simple
explanation using the model, Fig. 2.

When TMA/Methanol interacts with DNA sequence 1 with 21 bases, which
consists of 5 Thymine, 10 Guanine, 2 Cytosine and 4 Adenine, each
Gas-DNA-base complex gives $\alpha$, $\beta$, $\gamma$ \& $\delta$
negative ions that causes a net charge transfer from the
Gas-DNA-base complex to the SWCN. Similarly when TMA/Methanol
interacts with sequence 2 with 24 bases, which consists of 11T's,
4G's, 5C's \& 4A's, the Gas-DNA-Thymine complex gives $6 \alpha^{-}$
ions more than the value of $\alpha^{-}$ ions we get with sequence
1, whereas Gas-DNA-Guanine complex gives $6 \beta^{-}$ ions less
than sequence 1, Gas-DNA-Cytosine complex gives $3 \gamma^{-}$ ions
more than sequence 1 and Gas-DNA-Adenine complex gives $4
\delta^{-}$ ions, which is the same as in sequence 1. We propose
that when TMA/Methanol interacts with sequence 1 (2), the net charge
transfer is dominated by transfer of $\alpha^{-}$ ions (the value of
$\alpha>\beta,\gamma,\delta$) which is more for sequence 2 than
sequence 1 (because sequence 2 has 11T's, but sequence 1 has only
5T's each giving $\alpha^{-}$ ions). Hence for TMA/Methanol, the
current decreases and confirms the result shown in Table 1
\cite{17}. The value of $\alpha^{-}$ ions is different for TMA and
Methanol because both have different current responses. For TMA the
decrease in the current is large because of an additional charge
transfer due to direct interaction of TMA with the SWCN.

On the other hand, there is increase in the current when DMMP/DNT
interacts with sequence 2 as compared with sequence 1. This
indicates that the net charge transfer from the Gas-DNA-base complex
to the SWCN is increased as we go from sequence $2$ to $1$. For
DMMP/DNT the net charge transfer is dominated by transfer of
$\beta^{-}$ ions (with $\beta>\alpha,\gamma,\delta$) which is small
for sequence 2 as compared with sequence 1. The interaction of
DMMP/DNT with 10G's of sequence 1 causes the Gas-DNA-Guanine complex
to transfer more $\beta^{-}$ ions than the $\beta^{-}$ ions
transferred by the interaction with sequence 2 (with 4G's).

For PA there is decrease in the current when PA interacts with
sequence 2 than when it interacts with sequence 1. This shows that
the net charge transfer from SWCN to the Gas-DNA-base complex is
low. Again the net charge transfer is dominated by transfer of
$\beta^{-}$ ions ($\beta > \alpha,\gamma,\delta$). Hence we see
that the odor response characteristics of these sensors are
specific to the ss-DNA sequences.

It follows from the model that two distinct sequences with the same
number of bases but different order of the bases will give the same
minimum value of the current with different current characteristics
(active part).

\section{Results} \label{5.}
\subsection{Autocorrelation function}

The expression for the autocorrelation function is

\begin{equation}
P_k =  { {{1\over {N}}
\sum_{t=k}^{N-(k+1)}(I(t)-\bar{I})(I(t+k)-\bar{I}) }\over
   {{1\over {N}} \sum_{t=0}^{N-1}(I(t)-\bar{I})^2}}
\end{equation}

where $I(t)=I(t)/I_0$ is the normalized source-drain current at time
$t$ with $I_0$ the value of the current without odor, and
$\bar{I}={1\over N}\sum_{t=0}^{N-1} I(t)$ is the mean of the current
over the time interval $0\le t \le (N-1)$. $N$ is the total number
of observations and $k=0,1,2,3..$ is the time lag.

Data analytical methods used in the calculation of the
autocorrelation function may be helpful in realizing the sensitivity
and selectivity of a sensor. The autocorrelation function is a
pattern recognition measure that is useful for finding the repeating
patterns in a signal and for detecting long range correlations. The
autocorrelation function is a correlation of a data set with itself:
the correlation is between two values of the same variable at time t
and t + k averaged over the whole signal. In the current
characteristics \cite{17} the pattern for both gas and air repeats
itself. In the calculation we divide the complete pattern \cite{17}
into different segments for the exposure time when gas replaces air
Figs. 3 (a), (b) and when air replaces gas Figs. 3 (c), (d). For
each data segment we calculate the autocorrelation function then we
take the average of all the segments and plot the averaged data with
respect to time lag, Fig. 3.

We observe that the autocorrelation function is different for
different odors and it is different for both air and odors. The
autocorrelation function for air has some structure and is
different from the function for randomly generated numbers. The
autocorrelation function for odor together with air form a
distinct pattern of the Gas/Air-DNA-SWCN system. Note that the
autocorrelation function for Methanol with sequence 2, Figs. 3
(a), (c) is different from Methanol with sequence 1, Figs. 3 (b),
(d). This shows that the autocorrelation function varies for the
same odor with different DNA sequences. Hence the autocorrelation
function gives distinct patterns of odor sensitivity indicating
the selectivity of a sensor to a particular target odor. Taking
logarithm (decade) of the autocorrelation function for Methanol
with sequence 1 \& 2 and plotting it with respect to time lag we
get different slopes for both gas and air, Fig. 4. The different
slopes indicate that the autocorrelation function is sensitive to
the DNA sequence and strengthens the results found in Fig. 3. The
small differences in the slopes show high sensitivity of the
sensor. For other odors also we find different slopes indicating
the ability of ss-DNA/SWCN-FET sensors to detect minor changes in
the current response for different odors.

Our aim in calculating the autocorrelation function is to get a
pattern of close points for different odors and sequences, which may
relate the ss-DNA sequence with the sensor response. This long range
autocorrelation function is sensitive to the full DNA sequence.
Using the data from the current characteristics \cite{17} we
calculate the autocorrelation function using equation 1.

\subsection{Two-point correlation function}

Let us define the two-point correlation function $G(t)$ for the
current as

\begin{equation}
G(t) =  { {1\over \tau} \sum_{i=1}^{\tau}I_{i}(t)I_{i}(t+\delta
t)-\left({1\over \tau}\sum_{i=1}^{\tau}I_{i}(t)\right)^2}
\end{equation}

where $I_{i}(t)=I_{i}(t)/I_0$ is the normalized source-drain current
in the $i^{th}$ gas/air exposure cycle, $\delta t = 1$ s  and $\tau$
is the the total number of exposure cycles.

Figs. 5 \& 6 show the results of this calculation for different
odors and DNA sequences upon odor and air exposures. Though there
are a lot of fluctuations in the two-point correlation we still
observe distinct patterns for the different odors and sequences.
This indicates selective recognition of each odor by the sensors.
Table 1 shows that the values of correlation function $G(t)$ upon
different odor and air exposures lie in different ranges.

Note the distinct patterns for Methanol with sequence 2 and 1, Figs.
5 (c), (f) and 6 (c), (f). This shows that the Figs. 5 and 6 may be
used along with the Fig. 3 to identify the odor and DNA sequence.

\section{Discussion and Conclusion}
\label{6.}

We studied the experiment \cite{17} which involves a change in
sensor current upon odor exposure relative to the level measured
when the sample is exposed to clean air. The sensor response curves
\cite{17} show that the bare SWCN-FET device does not respond to any
odor except TMA. We propose that the odors can interact with the
SWCN through ss-DNA and the interaction could be the vdW forces
and/or mutual polarization \cite{16,20} between the odors and
DNA-SWCN. The model (Fig. 2) explains that the interaction between
the gas and DNA causes charge redistribution and changes the
polarization between the DNA and the SWCN leading to a net charge
transfer from the Gas-DNA-base complex to the SWCN and vice-versa
for PA. This is responsible for the sensor response.

For PA, the gas molecules interact with the full DNA sequence
before completion of the exposure time (let this time be $t_c$)
therefore we find a positive value of the current which remains
constant throughout the odor exposure after $t_c$. This shows
flatness (saturation) in the pattern of the current
characteristics. For methanol with sequence 1 a similar flatness
is observed for the same reason.

In experiment \cite{17} we find that to maintain a constant sensor
response the odor concentration should be constant. Usually, e.g.
in Ref. \cite{9}, the sensor response is linear for different
concentrations but in experiment \cite{17} this is not observed as
the current is a function of time for a fixed concentration. To
detect the sensor response the odor concentration and flow rates
should be enough to cause charge transfer which is necessary for
the sensing material to sense the target odor and should be such
that the molecules interact with the full DNA sequence to get the
repeated pattern. If the concentration is very low and not enough
to cause charge transfer then no response will be observed. If the
concentration is low but enough to cause charge transfer, the gas
will not occupy the full DNA sequence and the pattern will not
repeat itself. If the concentration is higher than required to
occupy the full DNA sequence then flatness will occur which
limits the characteristic part (active region) of the pattern.

In summary, we have explained the difference in the odor response
characteristics of the sensors for two different sequences using
the model and calculated the correlation functions indicating
that CNTs can be used as chemical sensors as well as sequence
detectors. We find from Figs. 3, 5 and 6 that the fluctuations in
the current of CNTs are more sensitive to the adsorption of
certain types of gases. The correlation plots of ss-DNA/SWCN-FET
sensors are different for different odors. Here the
autocorrelation function clearly shows that the Methanol with
sequence 2 is different from Methanol with sequence 1, Fig. 3.
The slopes for Methanol and air with sequence 1 (2) are different
(Fig. 4) indicating that the autocorrelation function is
sensitive to the DNA sequences and also show selectivity and
sensitivity of the sensor. The patterns for the two-point
correlators Figs. 5, 6 also show characteristic features of the
DNA sequence and odor used. These figures may also be used along
with the autocorrelation figures to identify the odor and DNA
sequences. The two-point current correlation function $G(t)$ has
a universal part and a non-universal (noise) part in mesoscopic
systems \cite{23}. The correlators that we have calculated here
may also have this structure. We are in the process of trying to
understand the physical meaning of these correlation functions
for the ss-DNA-Gas/SWCN complex systems in the context of the
tight binding model of CNTs.

\section* {Acknowledgements}

We would like to thank Professor A. K. Sood for encouragement. We
also thank Professors A. T. Charlie Johnson, Prabal Maiti, Umesh Waghmare
for discussions and J. C. Poler for critical comments. We would like to thank CSIR
project No. 03(1019)/05/EMR-II for financial support.

\section* {Biographies}
\textbf{Dr. N. Deo} received her Ph.D. degree in Physics in 1988
from Purdue University, USA after her post graduation in Physics
from University of Delhi, India. She held teaching and research
positions at Harvard University, Brown University, Tufts University,
Mount Holyoke College, Santa Fe Institute in the USA and the Indian
Institute of Science, Jawaharlal Nehru Centre for Advanced
Scientific Research, Raman Research Institute and PISR in Bangalore,
India. She was a Radcliffe Fellow at Harvard University in 1991-1992
and received the Edward Akeley Memorial Award for best graduate
student in Theoretical Physics in Purdue University, USA. She has
four Ph.D. students and $\sim$ 25 research publications with 10
single author publications. Her recent research interests are in
nanoscience, bioscience and complex systems.

\textbf{P. Poonam} received her B.Sc. degree and M.Sc. degree in
Physics from University of Delhi, Delhi, India in 2003 and 2005
respectively. She is currently pursuing the Ph.D in Delhi
University. Her current research interests are focused on the study
of functionalized nanomaterials by chemical and biological molecules
using numerical and analytical methods.

\newpage
\begin{figure}[h]
\includegraphics[width=12cm]{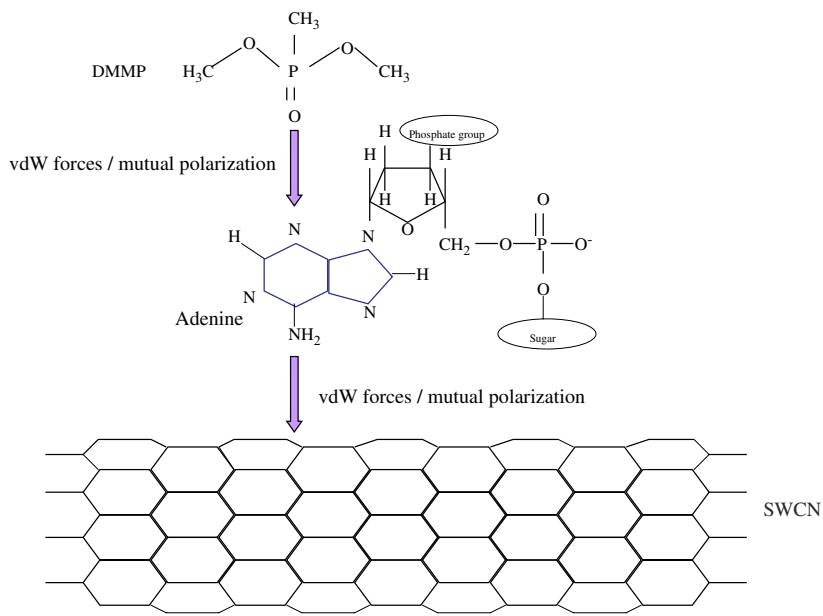}
\caption{\small{Schematic view of attachment of an adenine base of
the ss-DNA sequence 2 and a DMMP molecule on the SWCN through vdW
forces or mutual polarization.}}
\end{figure}

\newpage
\begin{figure}
\vspace{-4cm}
\includegraphics[width=11cm]{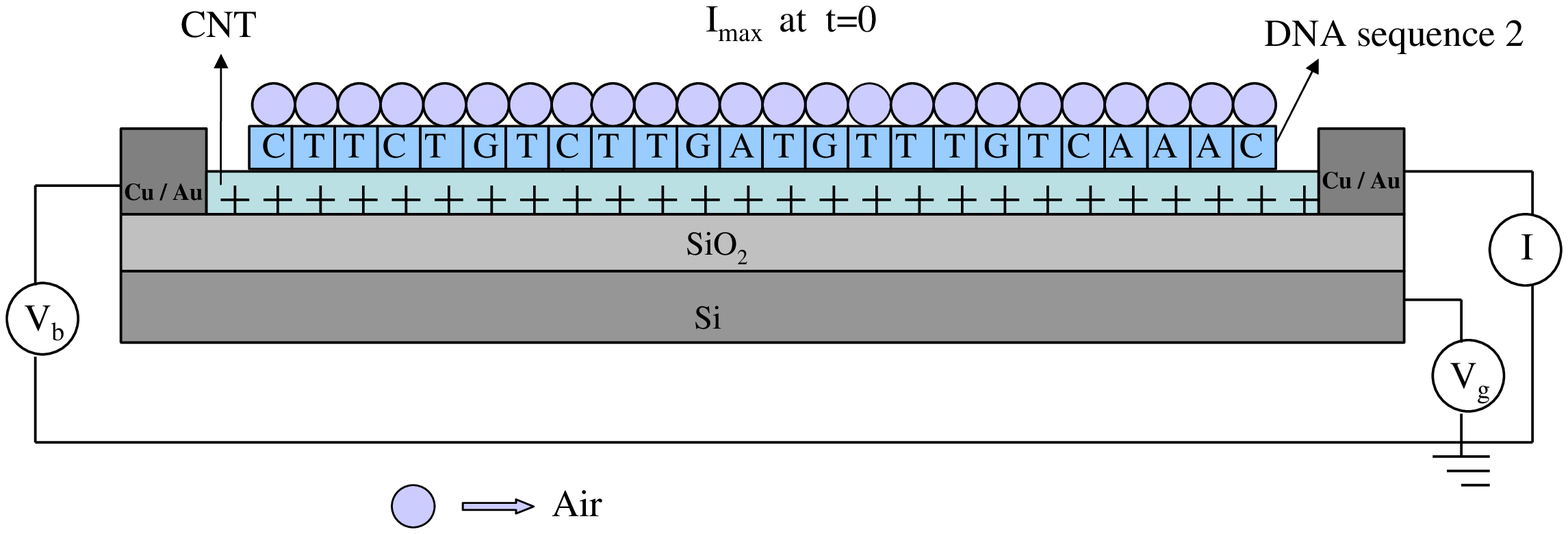}
\end{figure}
\begin{figure}
\vspace{-4cm}
\includegraphics[width=11cm]{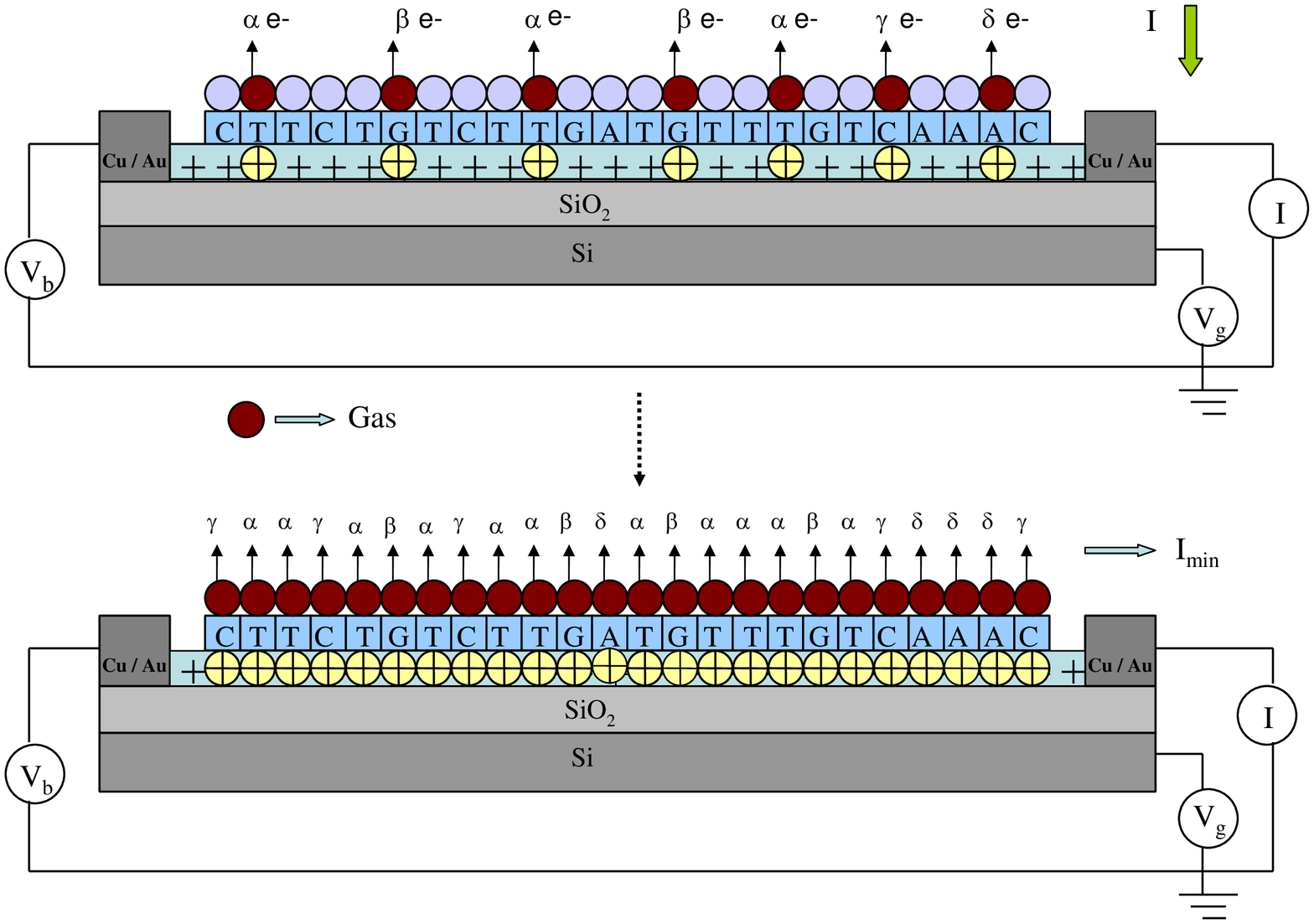}
\end{figure}
\begin{figure}
\vspace{-2cm}
\includegraphics[width=11cm]{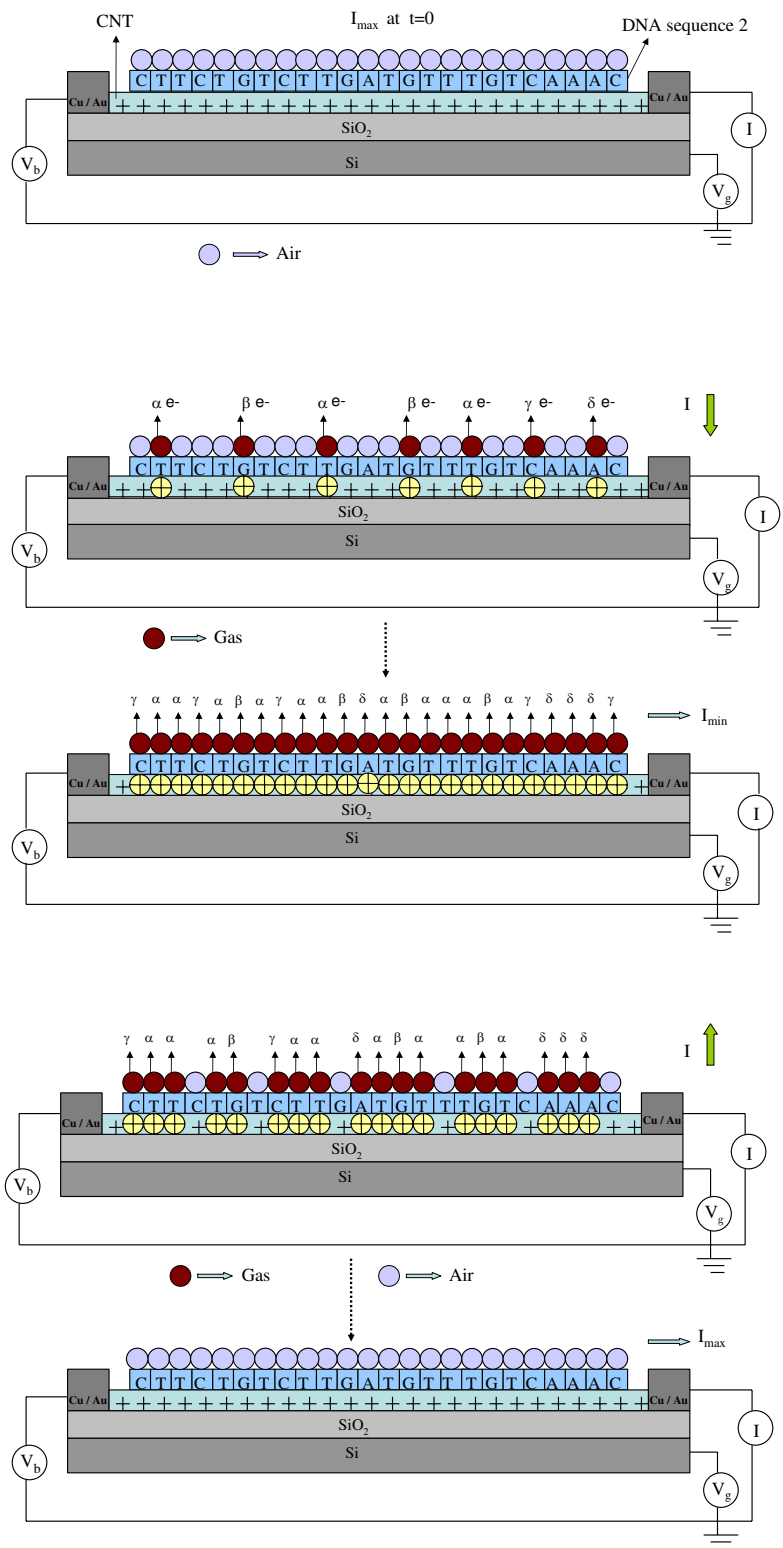}
\caption{\small{A phenomenological model. The brown and purple
balls indicate the gas and air, respectively.}}
\end{figure}

\begin{figure}
\includegraphics[width=7cm]{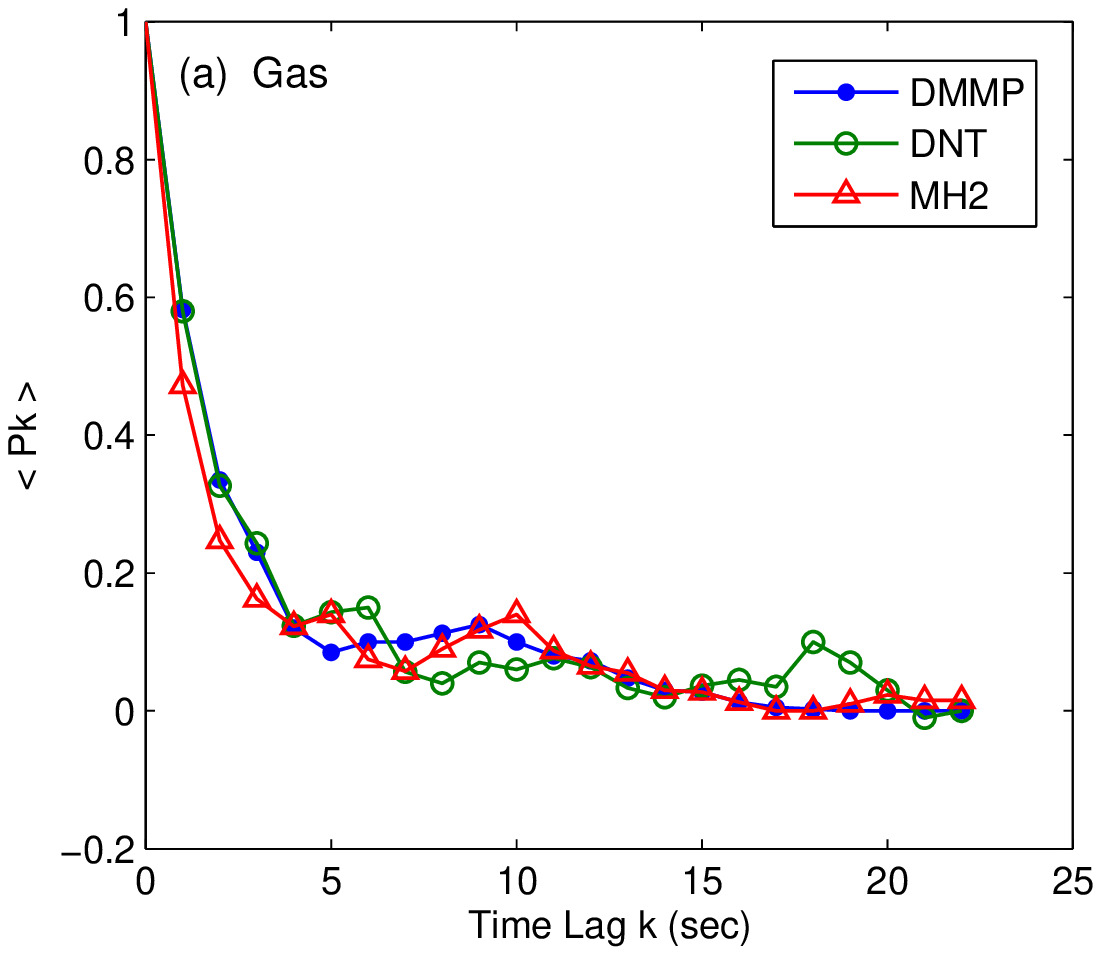}
\includegraphics[width=7cm]{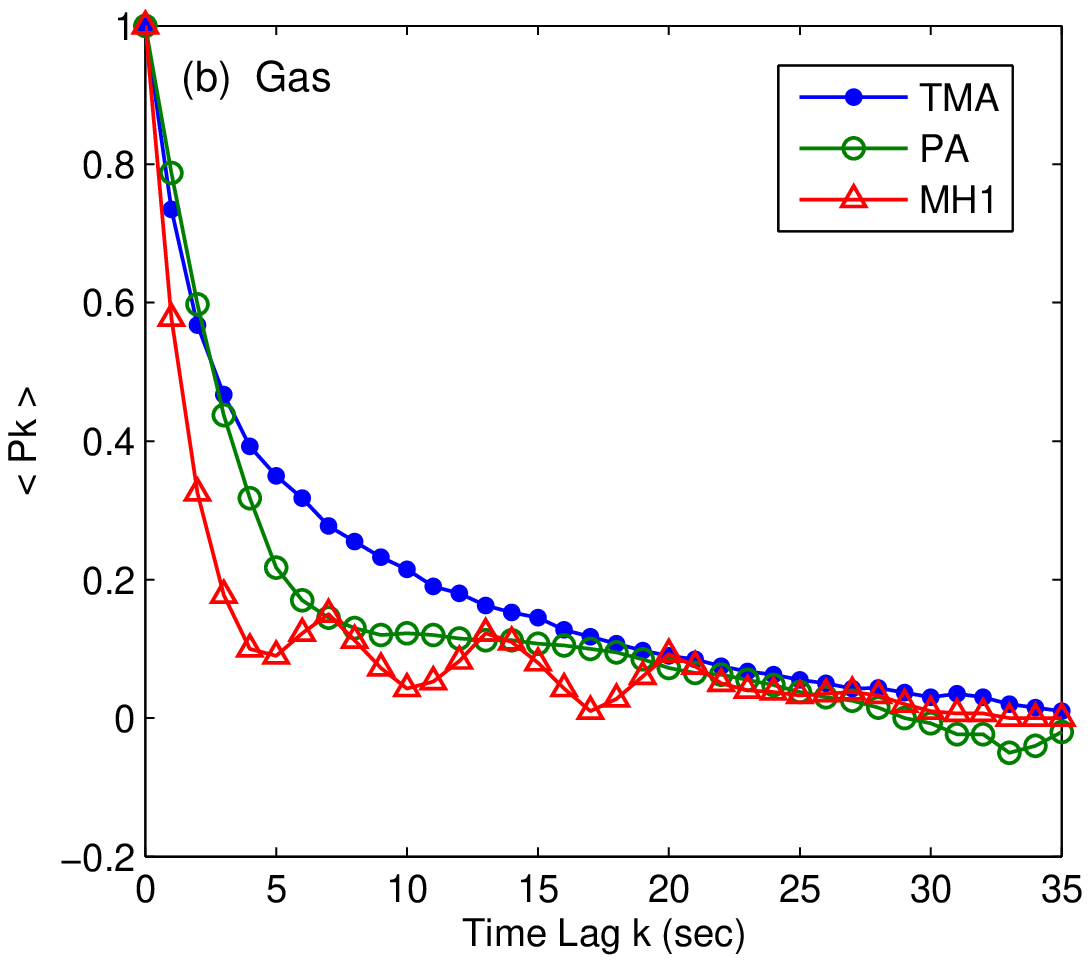}
\includegraphics[width=7cm]{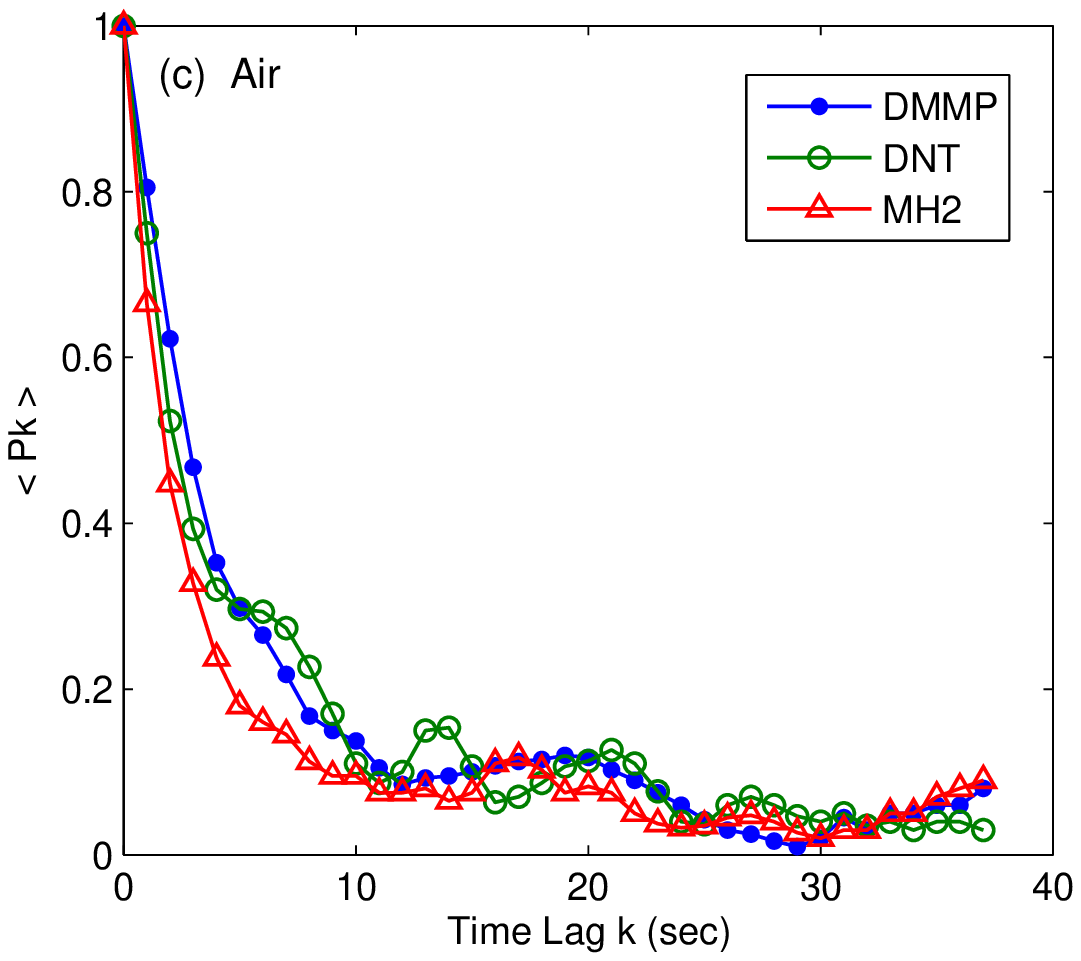}
\includegraphics[width=7cm]{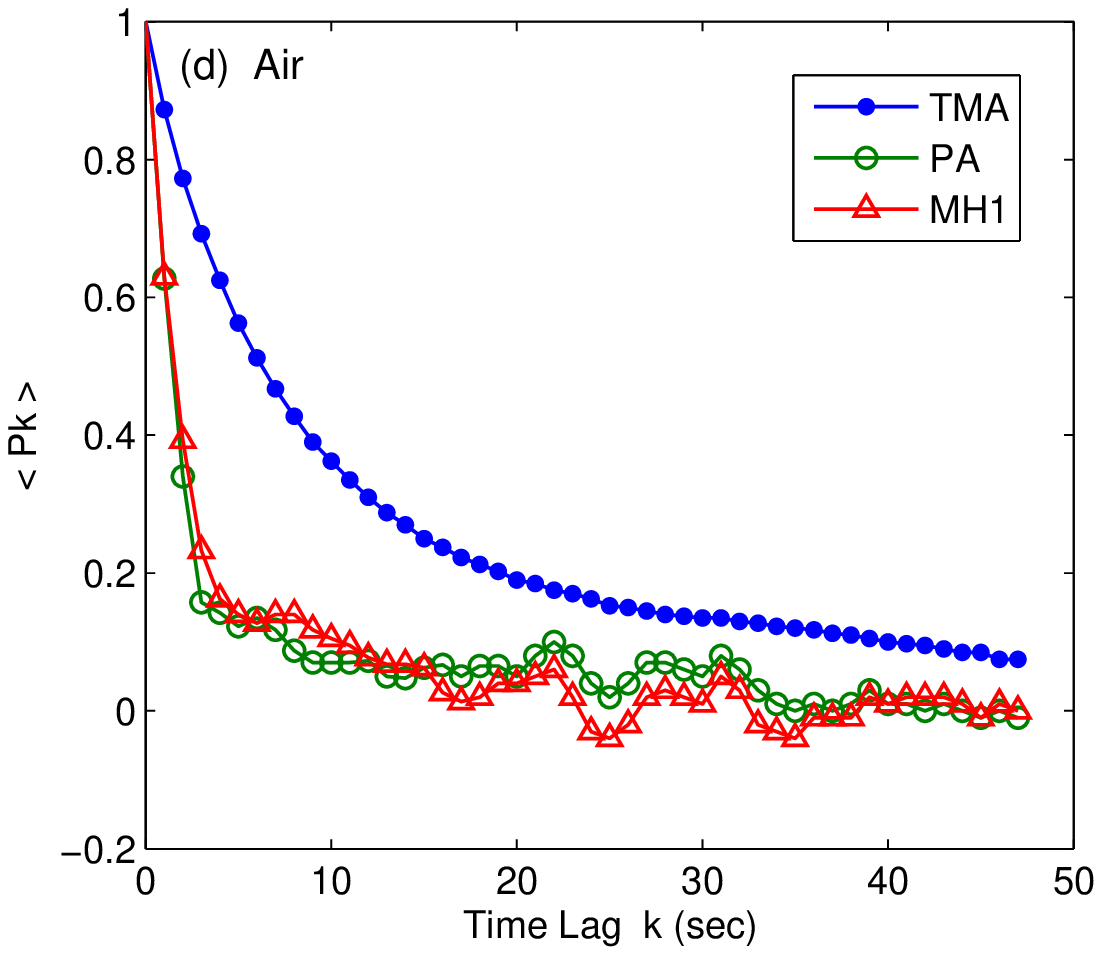}
\caption{\small{The autocorrelation function $<P_k>$ versus time
lag k(sec) for (a) DMMP + sequence 2, DNT + sequence 1 and
Methanol + sequence 2 upon gas exposure (b) TMA + sequence 2, PA
+ sequence 1 and Methanol + sequence 1 upon gas exposure (c) DMMP
+ sequence 2, DNT + sequence 1 and Methanol + sequence 2 upon air
exposure (d) TMA + sequence 2, PA + sequence 1 and Methanol +
sequence 1 upon air exposure. These plots distinguish between
different odors and DNA sequences.}}
\end{figure}

\begin{figure}
\includegraphics[width=7cm]{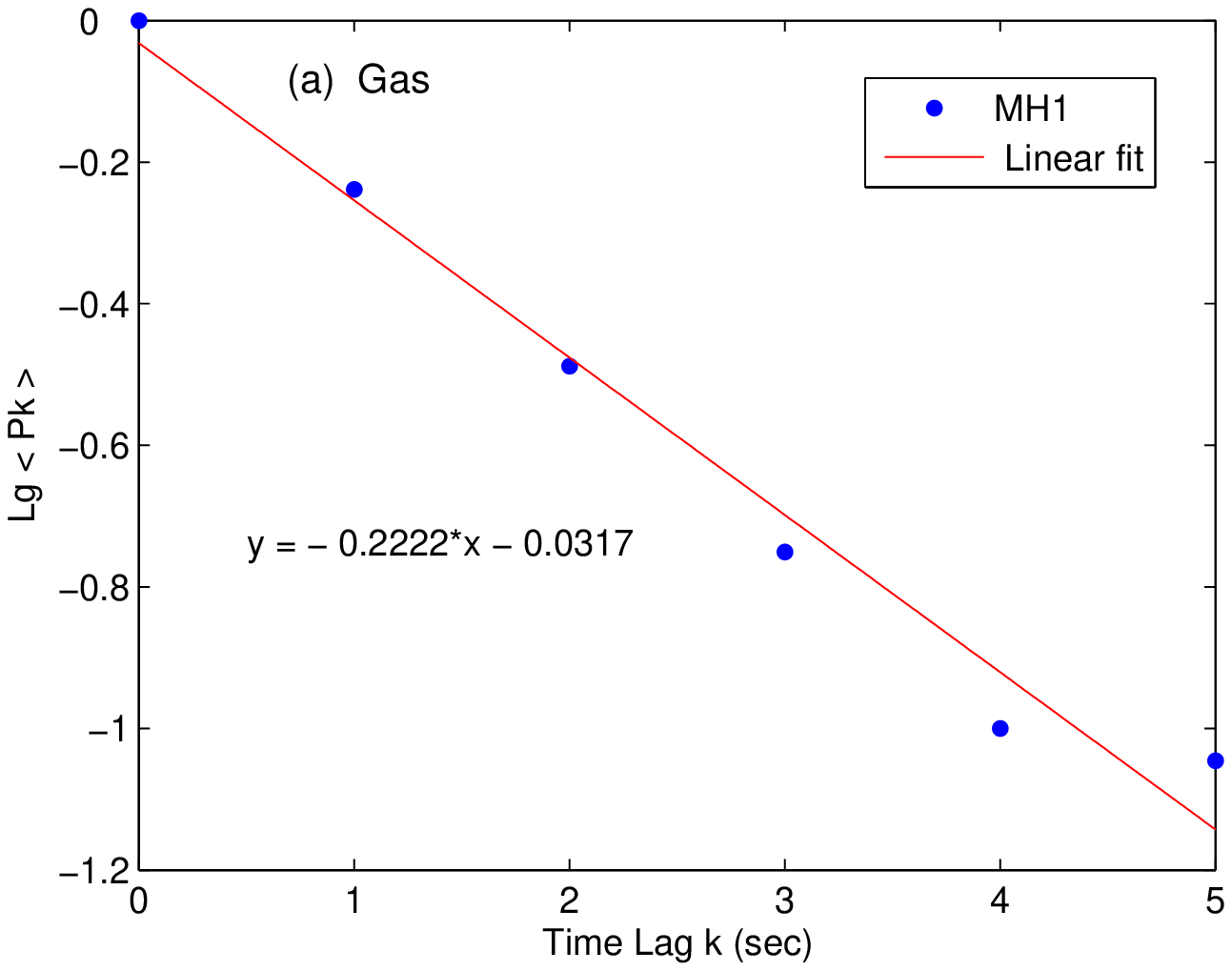}
\includegraphics[width=7cm]{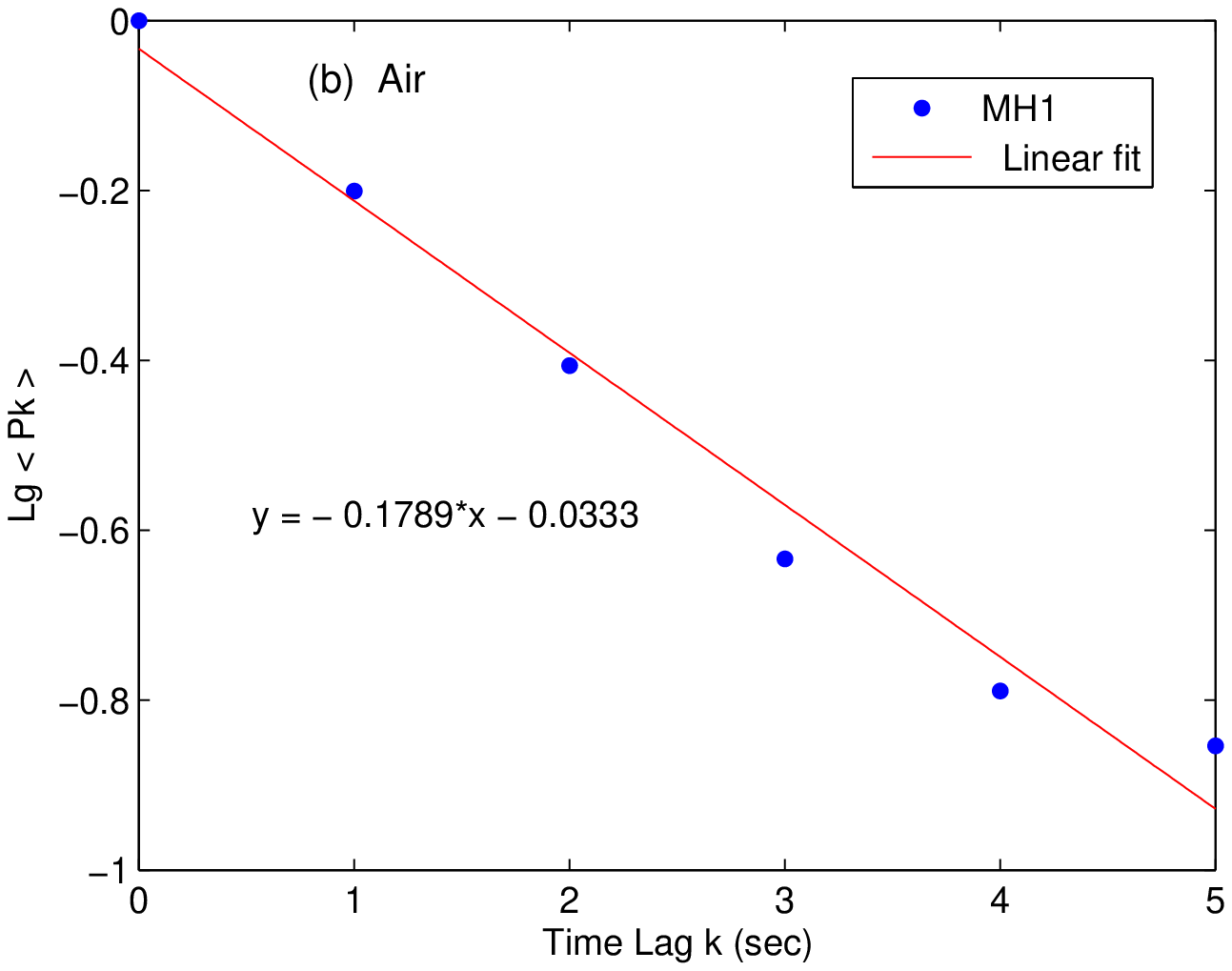}
\includegraphics[width=7cm]{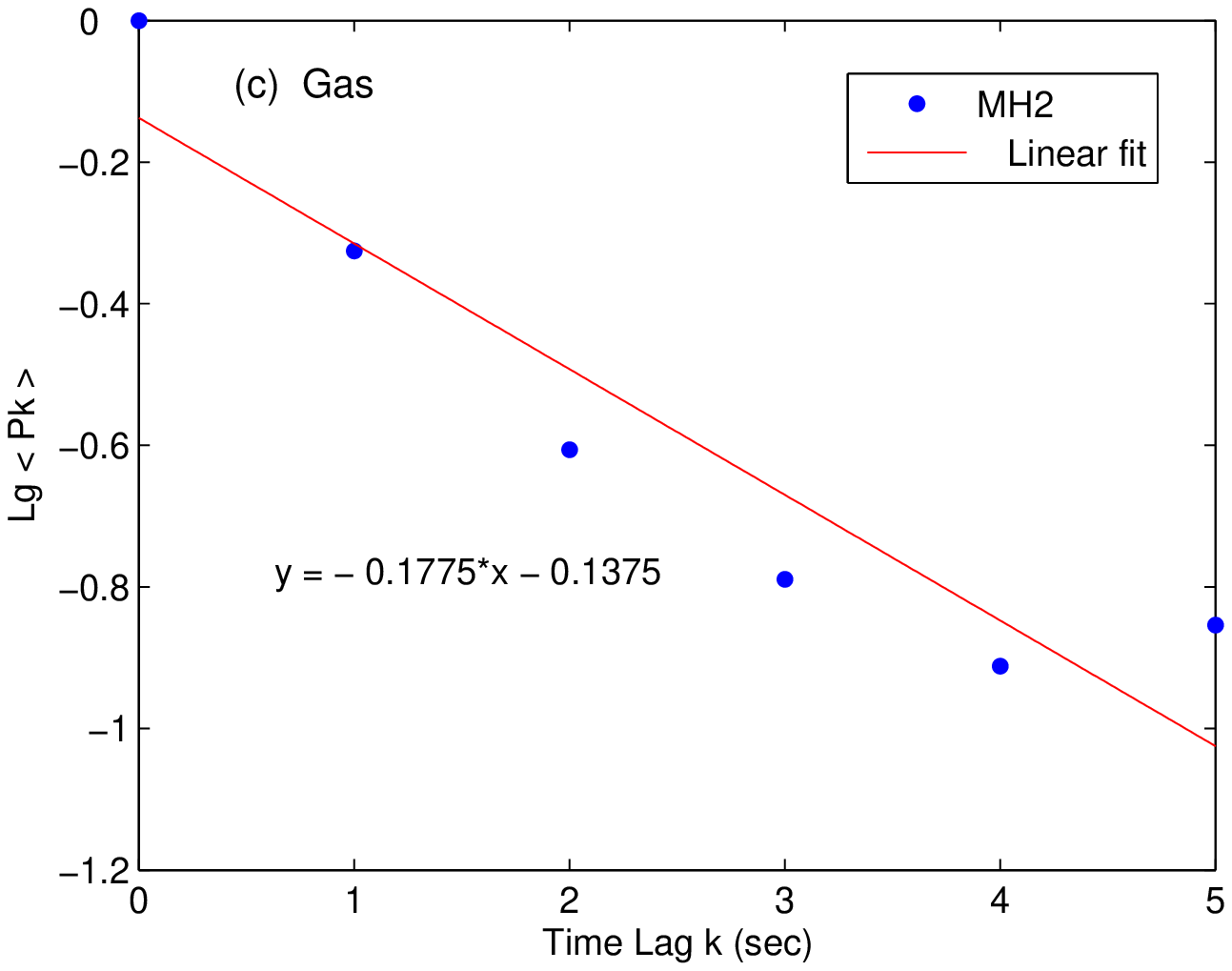}
\includegraphics[width=7cm]{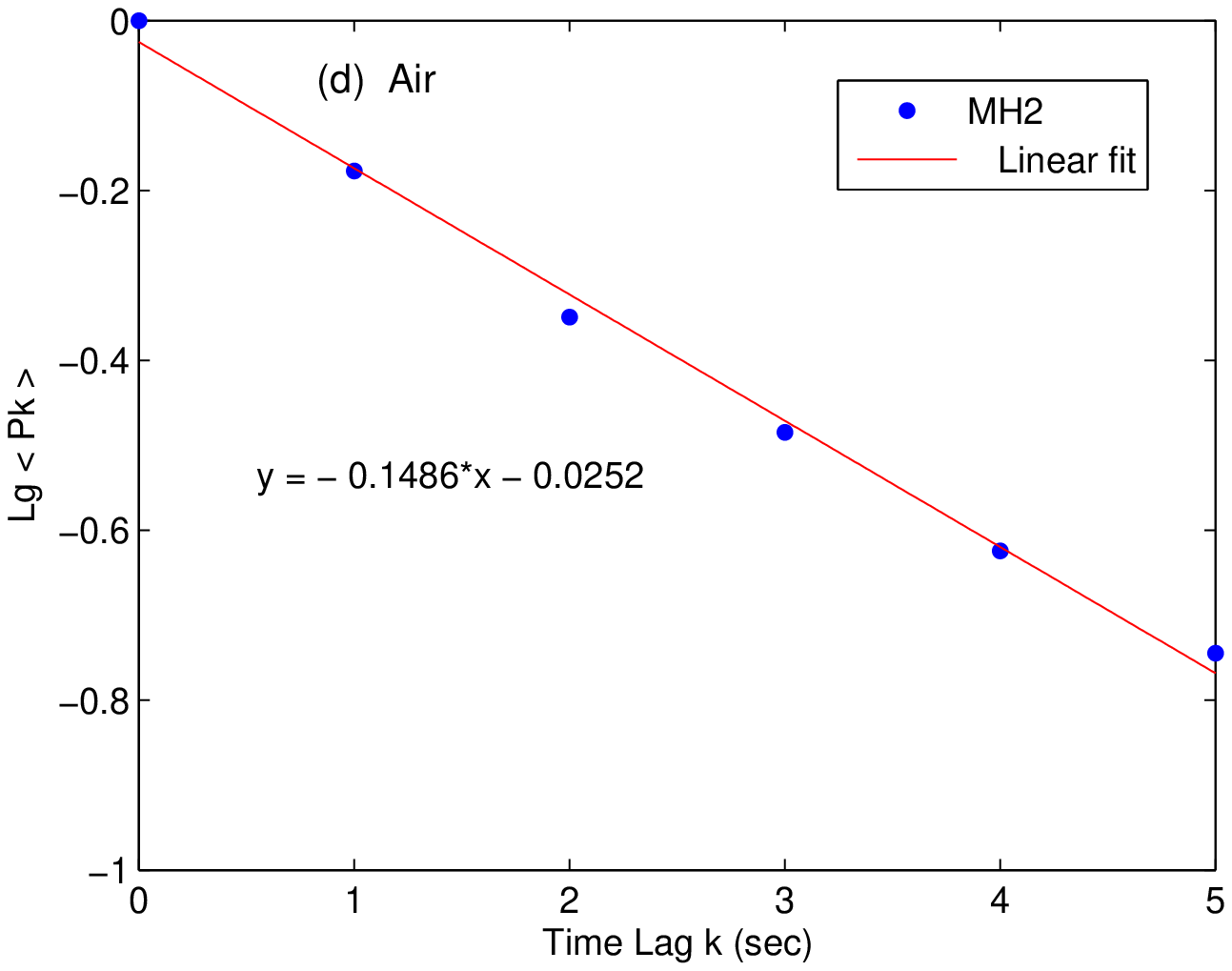}
\caption{\small{The decade logarithm of autocorrelation function
(y=lg$<P_k>$) versus time lag (x=k(sec)) upto 5 seconds for (a)
Methanol + sequence 1 upon gas exposure (b) Methanol + sequence 1
upon air exposure (c) Methanol + sequence 2 upon gas exposure (d)
Methanol + sequence 2 upon air exposure. Different slopes indicate
that the autocorrelation function is sensitive to the DNA sequence
for methanol and the differences in the slopes show sensitivity of
the sensor.}}
\end{figure}

\begin{figure}
\includegraphics[width=7cm]{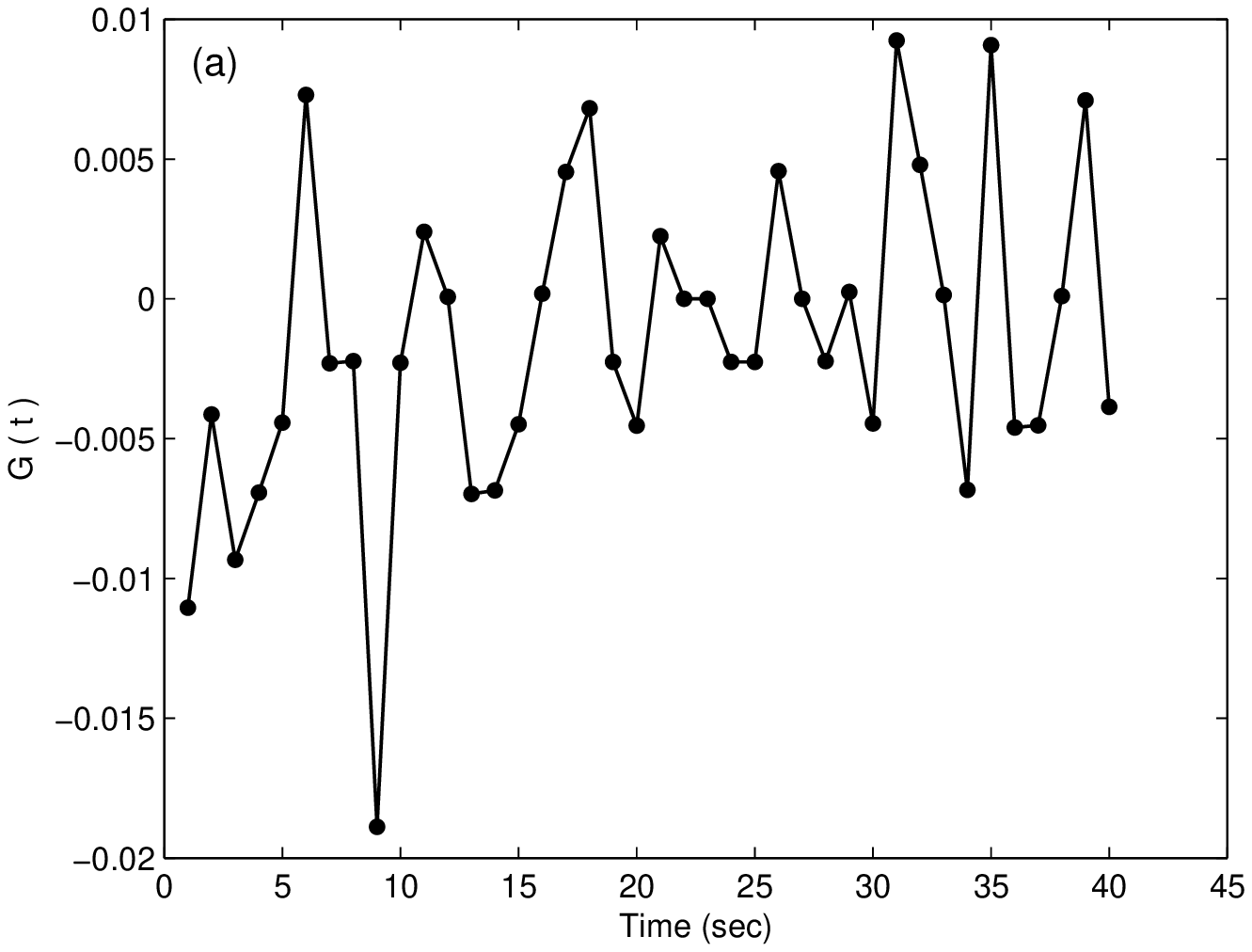}
\includegraphics[width=7cm]{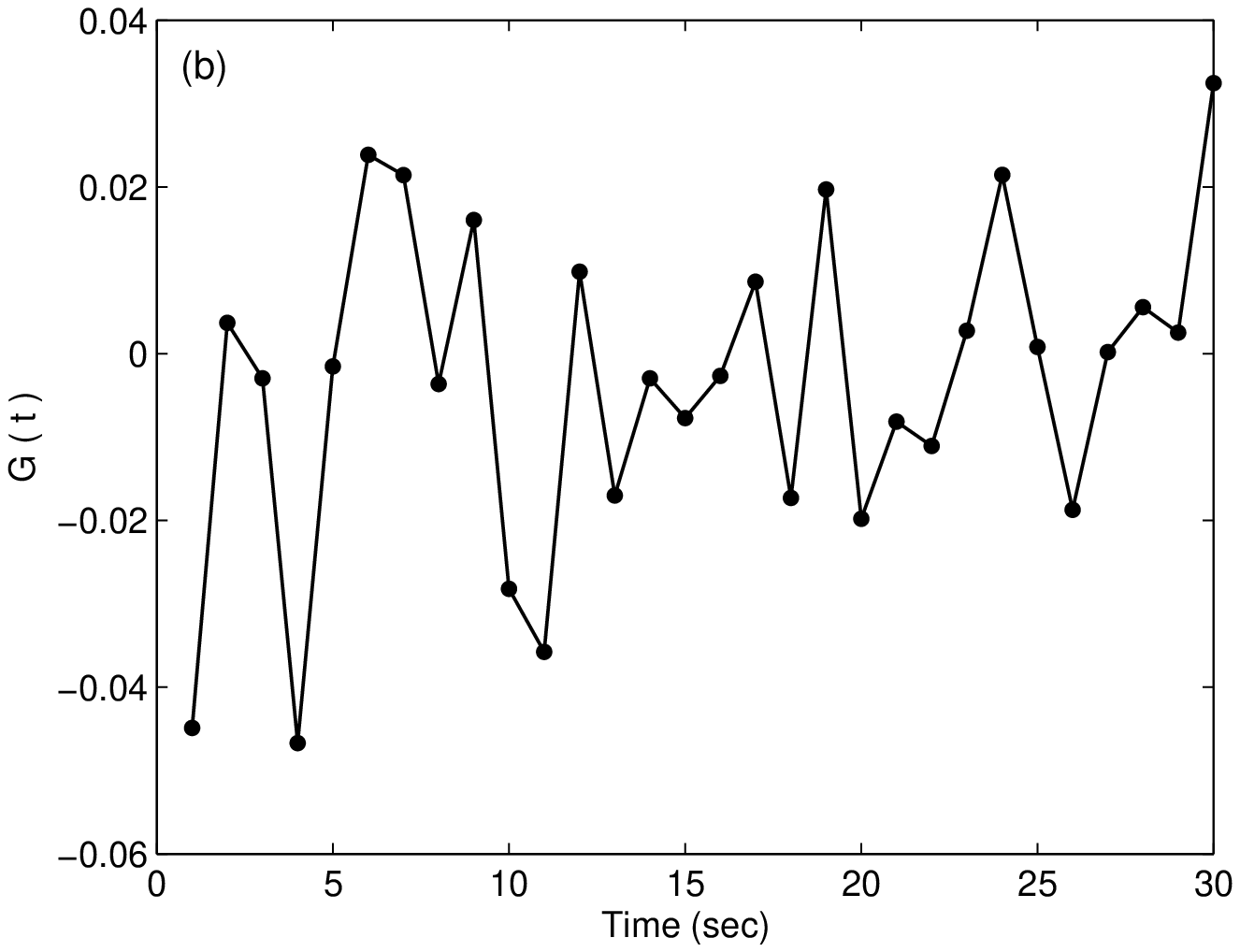}
\includegraphics[width=7cm]{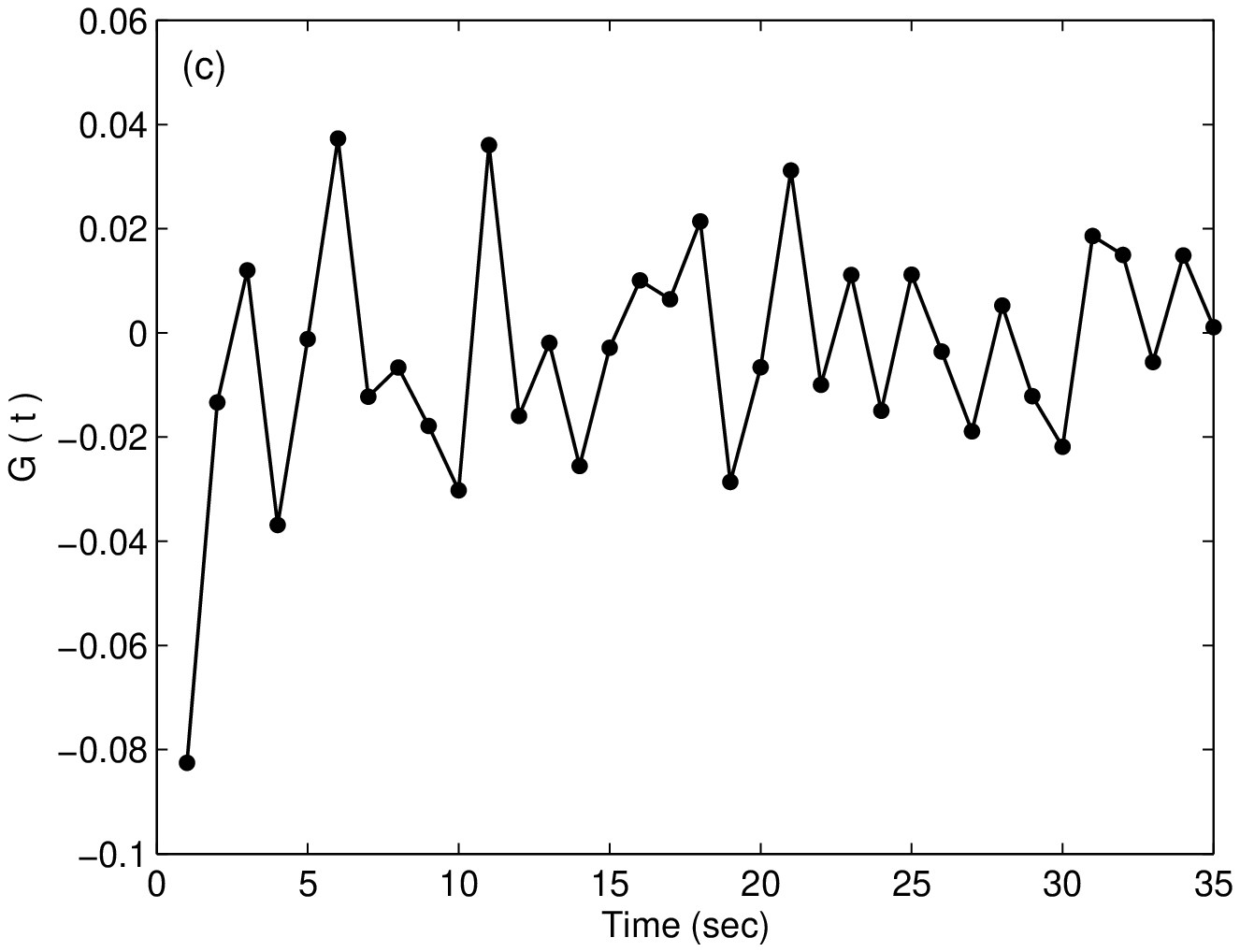}
\includegraphics[width=7cm]{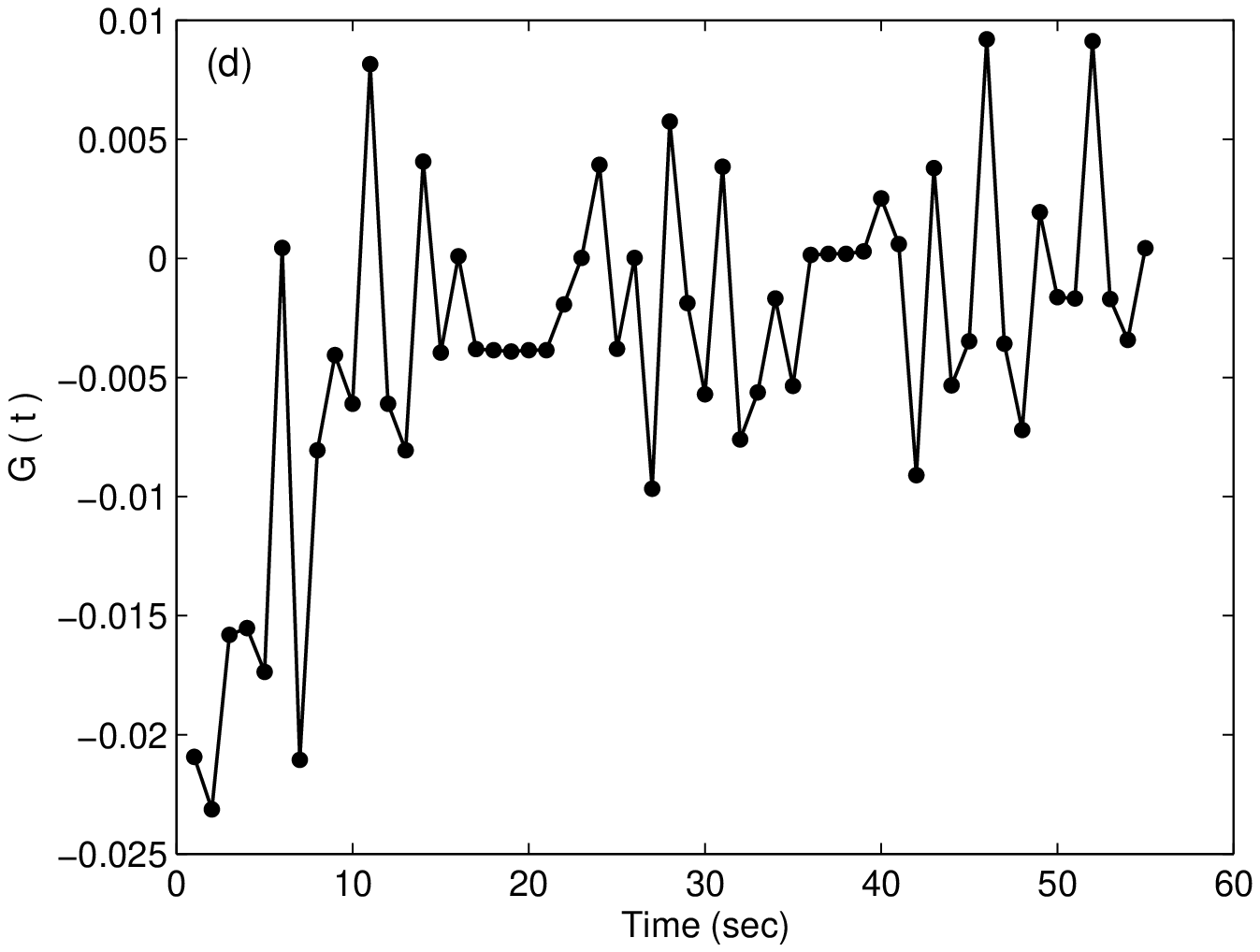}
\includegraphics[width=7cm]{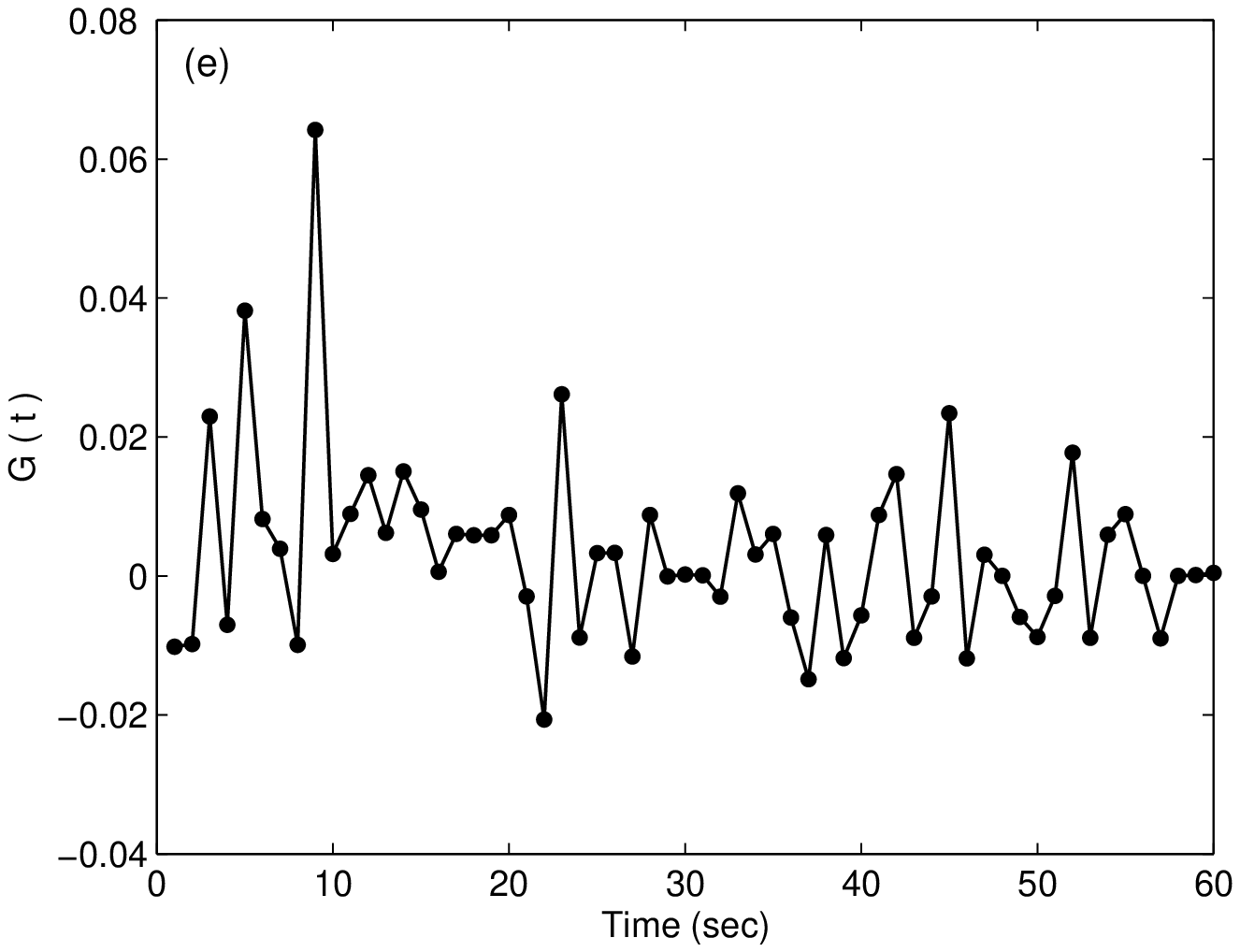}
\includegraphics[width=7cm]{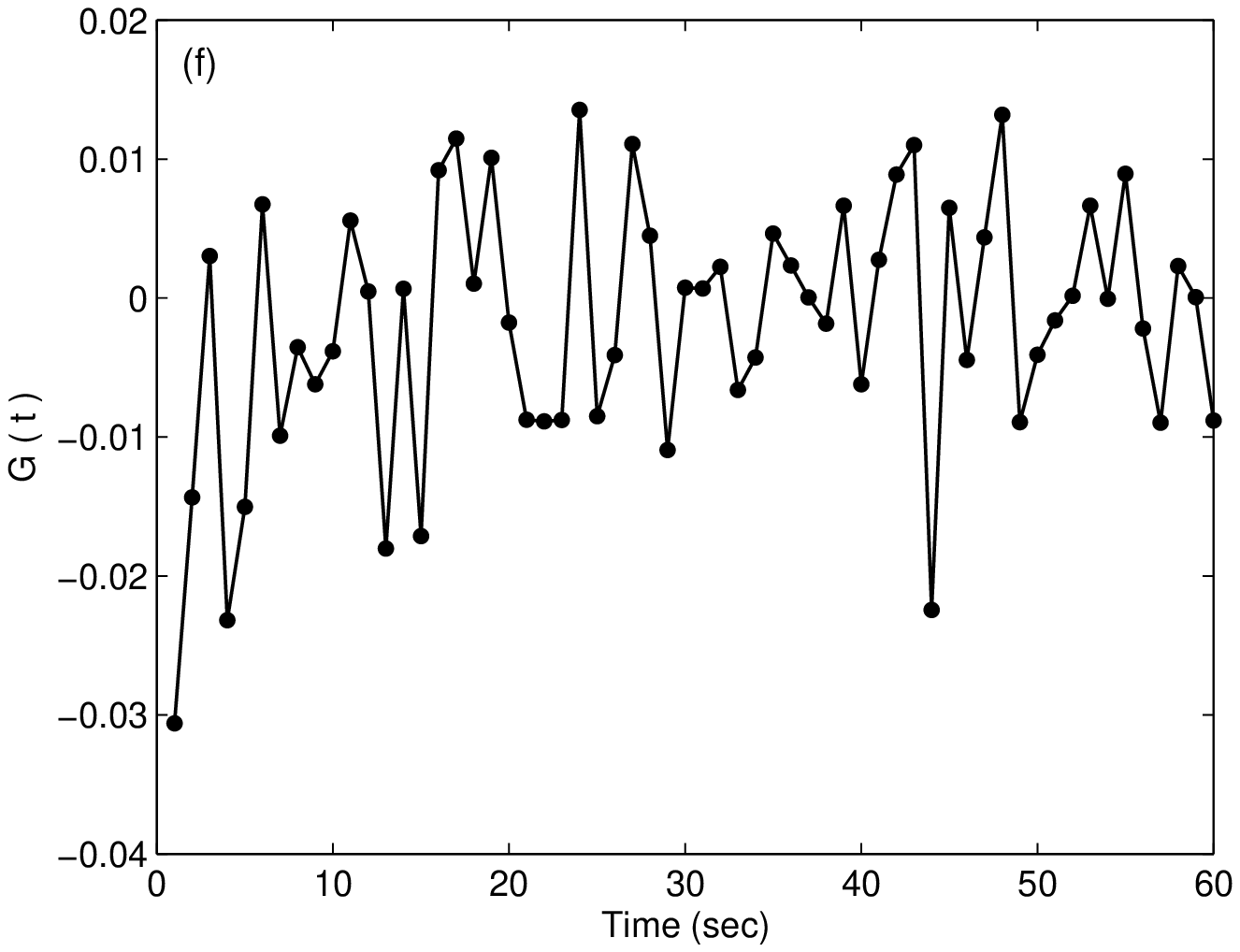}
\caption{\small{The two-point correlation function $G(t)$ versus
time(sec) upon gas exposure for (a) DMMP + sequence 2 (b) DNT +
sequence  1 (c) Methanol + sequence 2 (d) TMA + sequence 2 (e) PA
+ sequence 1 (f) Methanol + sequence 1. This indicates that the
fluctuations in the current of CNTs are more sensitive to the
adsorption of certain types of gases and DNA sequences.}}
\end{figure}

\begin{figure}
\includegraphics[width=7cm]{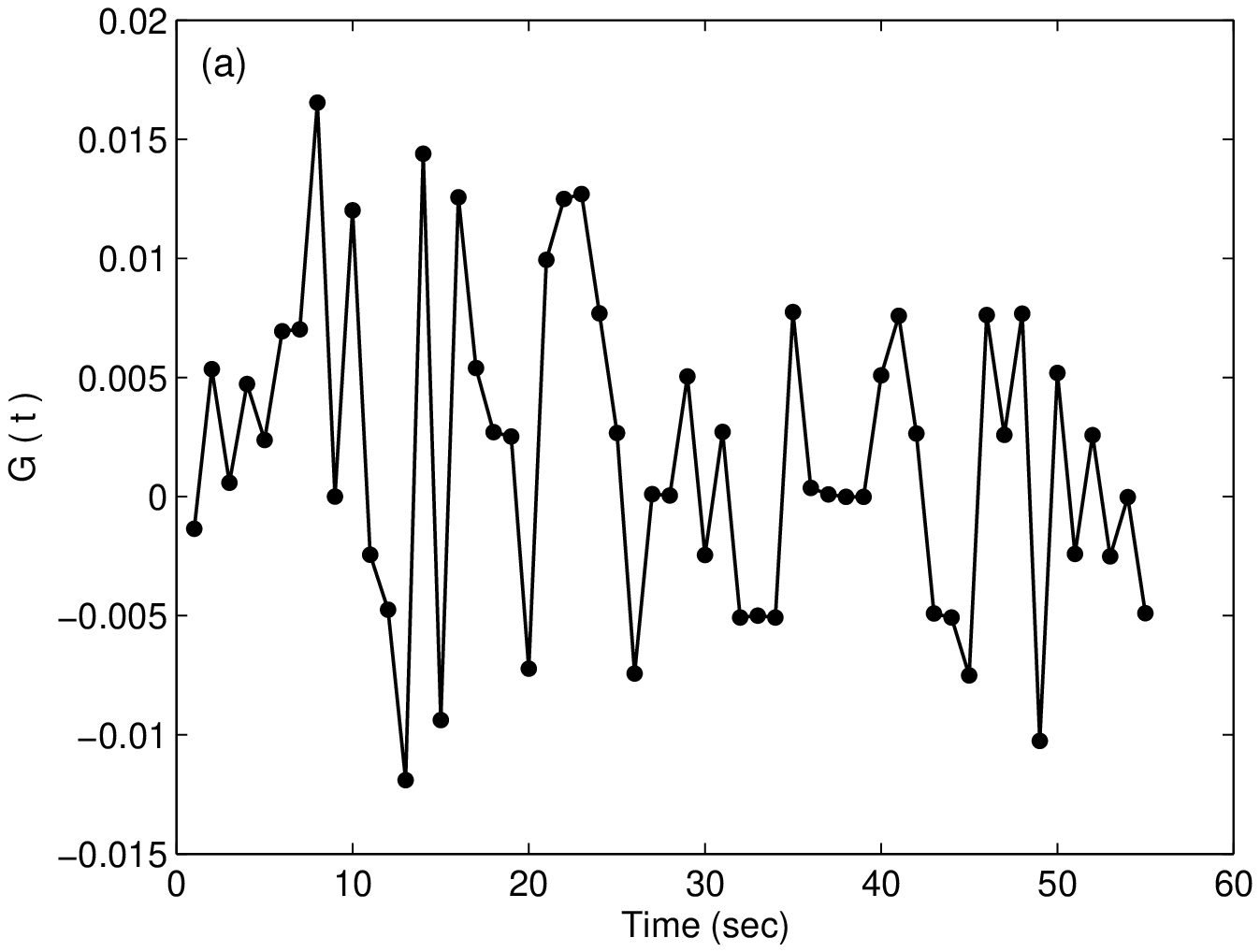}
\includegraphics[width=7cm]{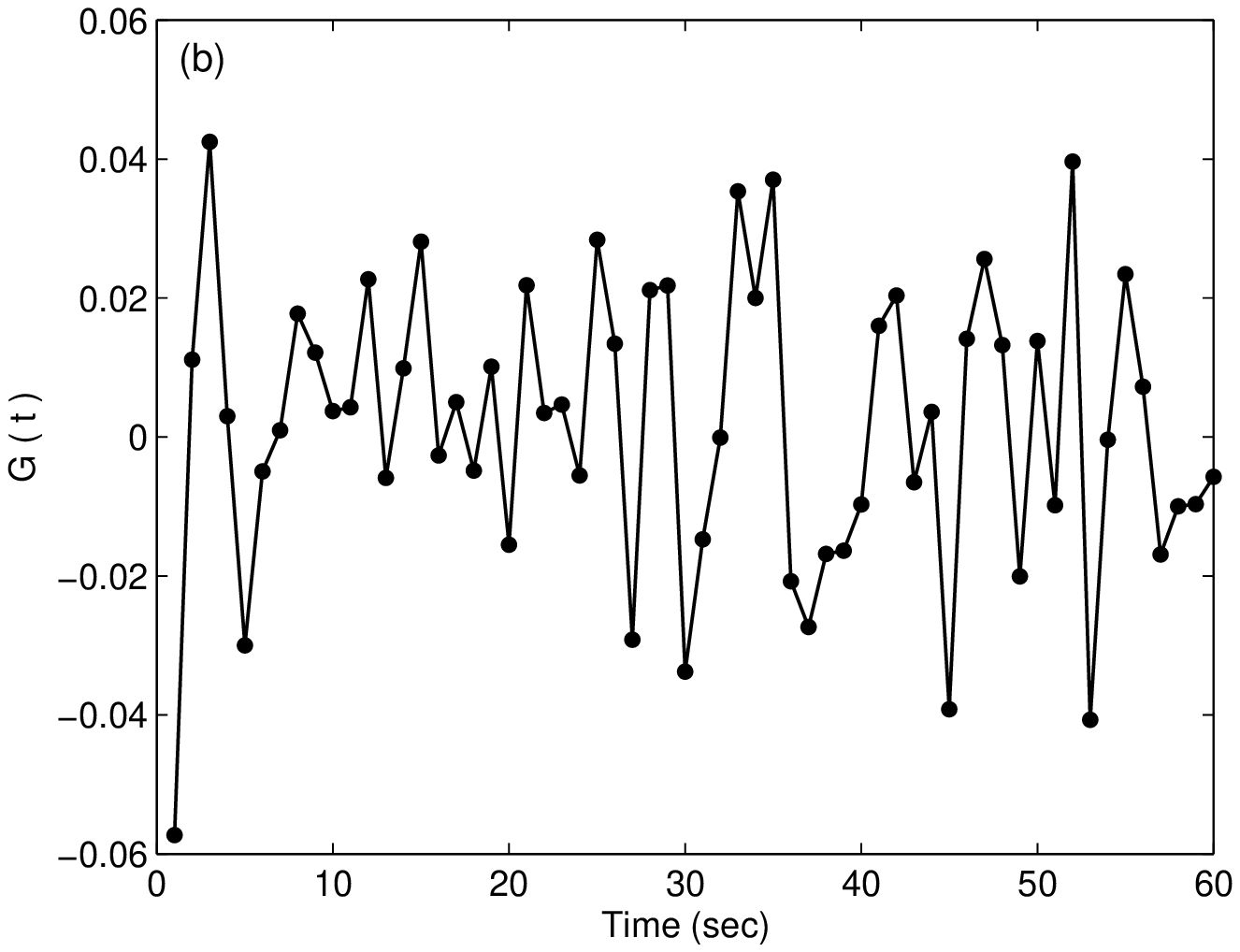}
\includegraphics[width=7cm]{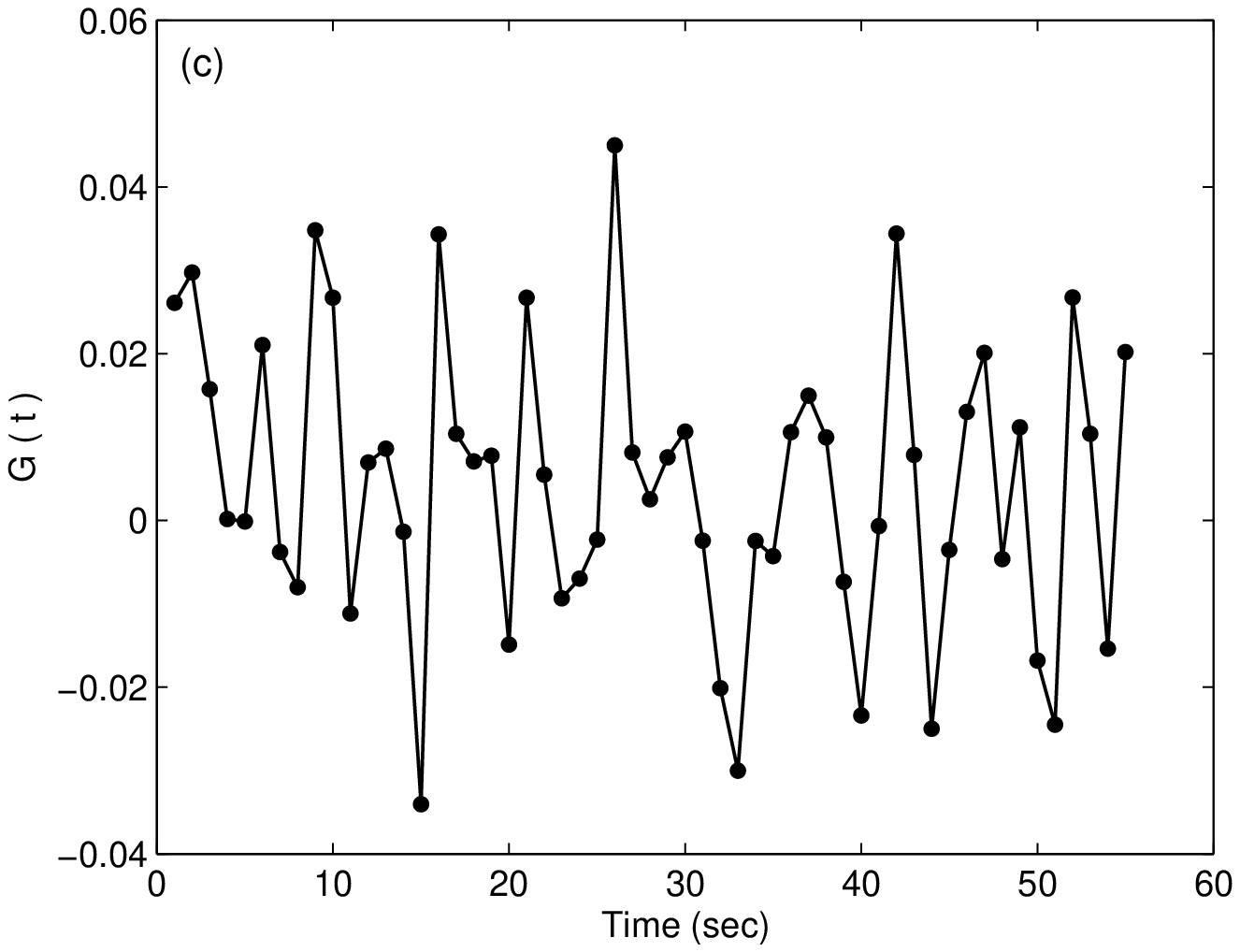}
\includegraphics[width=7cm]{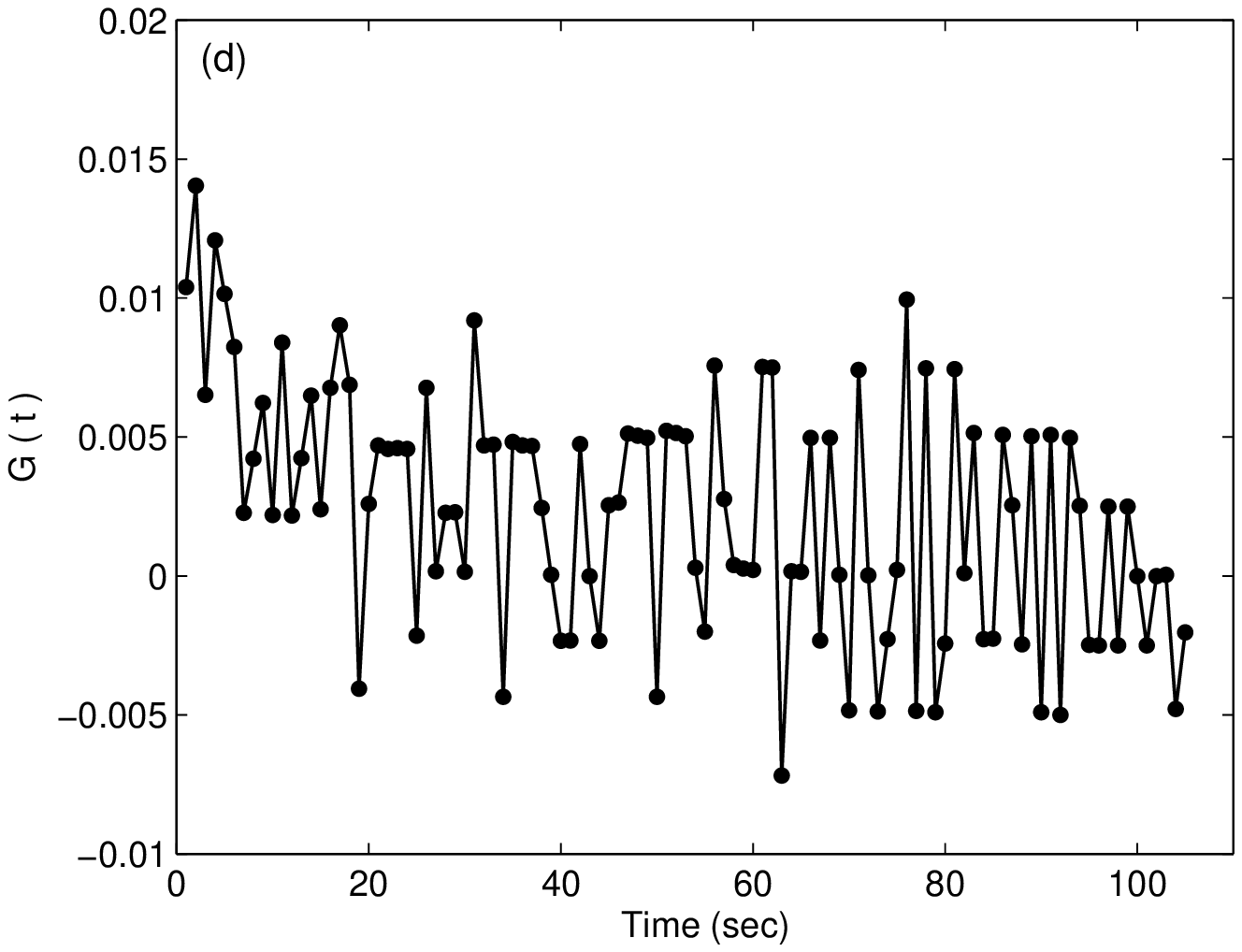}
\includegraphics[width=7cm]{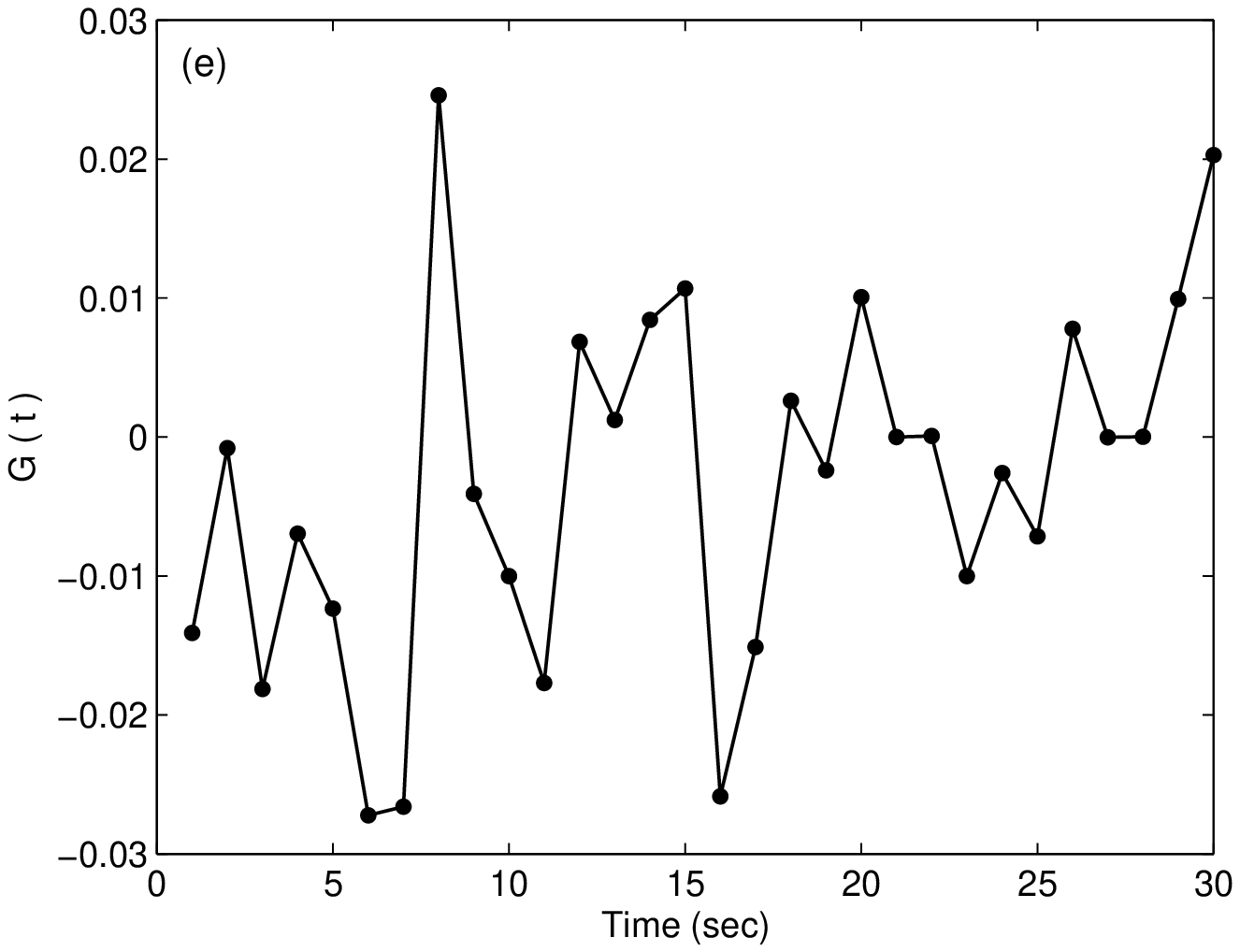}
\includegraphics[width=7cm]{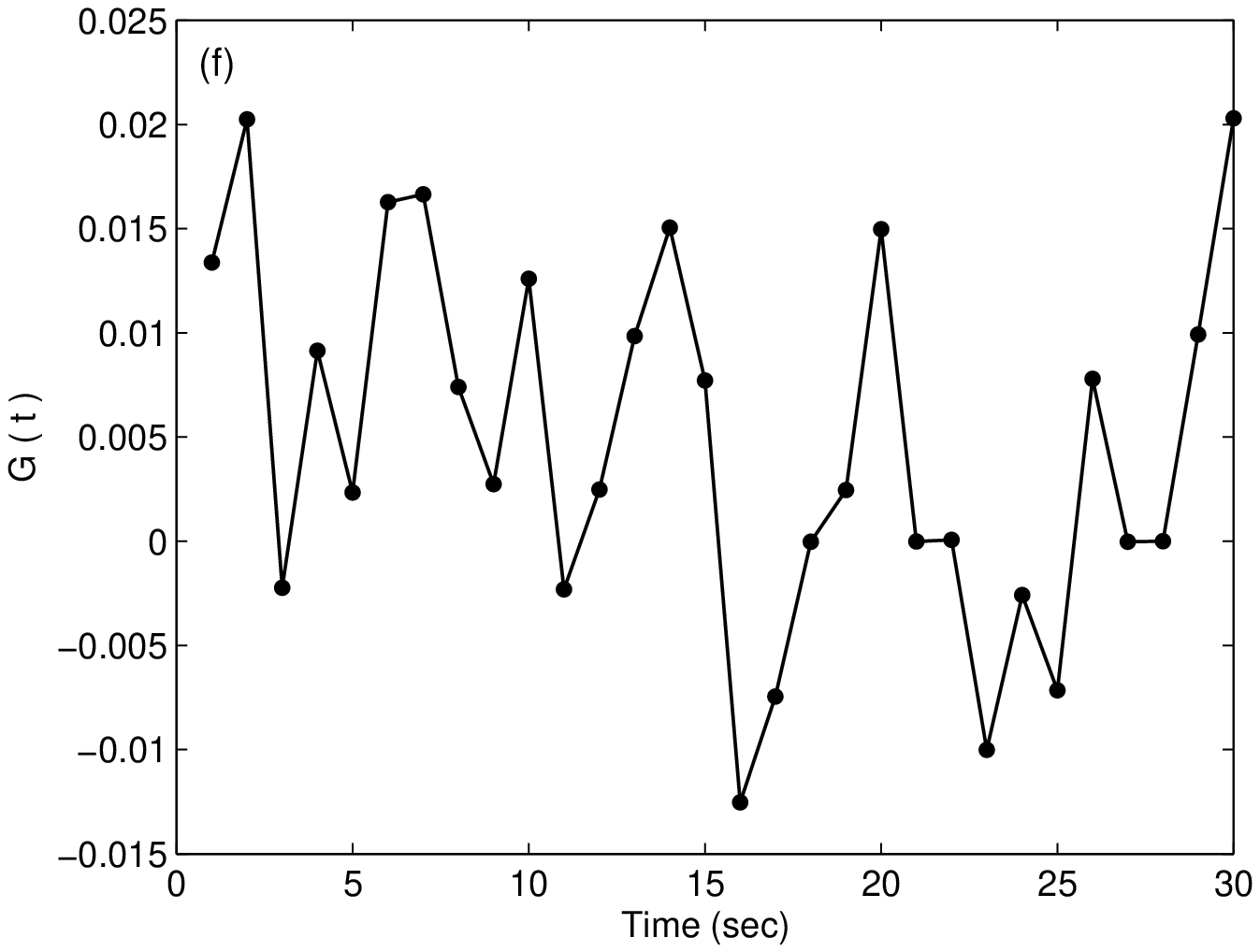}
\caption{\small{The two-point correlation function $G(t)$ versus
time(sec) upon air exposure for (a) DMMP + sequence 2  (b) DNT +
sequence 1 (c) Methanol + sequence 2 (d) TMA + sequence 2 (e) PA +
sequence 1 (f) Methanol + sequence 1. This indicates that the
fluctuations in the current of CNTs are more sensitive to the
adsorption of certain types of gases and DNA sequences.}}
\end{figure}

\newpage
\begin{table*}
\caption{Range of values of the two-point correlation function
$G(t)$ upon odor and air exposures} \vspace{0.5cm}
\begin{tabular}{lll}
\hline Odors \& sequence   & $G(t)\left(Odor\right)$   &
$G(t)\left(Air\right)$
\\

\hline
DMMP with sequence 2 & 0.009244 to -0.01888 & 0.016544 to -0.0119 \\

DNT with sequence 1 & 0.032467 to -0.04671 & 0.042489 to
-0.05728 \\

Methanol with sequence 2 & 0.037275 to -0.08255 & 0.044994
to -0.03406 \\

TMA with sequence 2 &  0.0092 to -0.02313 & 0.01405 to
-0.00718\\

PA with sequence 1 & 0.064194 to -0.02065 &  0.0246 to
-0.02723 \\

Methanol with sequence 1 & 0.01355 to -0.0306 & 0.0203 to
-0.01253\\
\hline
\end{tabular}
\end{table*}
\end{document}